\documentclass[preprint,numbers,12pt,fleqn,authoryear]{elsarticle}
\usepackage{booktabs,amsmath,amsthm,amssymb,amsbsy,latexsym,graphicx,color}
\usepackage{float}
\usepackage{geometry}
\usepackage{amsfonts}
\usepackage{rotating}
\usepackage{amsmath}
\usepackage{subfigure}
\usepackage{caption}
\usepackage{upgreek}
\usepackage{natbib}
\usepackage{bm}
\usepackage{mathdots}
\usepackage{relsize}
\usepackage[bottom]{footmisc}
\usepackage[bbgreekl]{mathbbol}
\newgeometry{vmargin={25mm}, hmargin={25mm,25mm}}

\journal{ArXiv}

\begin{document}
\begin{frontmatter}

\title{Finite strain constitutive modeling for shape memory alloys considering transformation-induced plasticity and two-way \\shape memory effect}

\author[ad1]{Lei Xu}
\ead{sdf007xulei@gmail.com}

\author[ad1]{Alexandros Solomou}

\author[ad3]{Theocharis Baxevanis}

\author[ad1,ad2]{Dimitris Lagoudas}

\cortext[bla]{Corresponding author}

\address[ad1]{Department of Aerospace Engineering, Texas A\&M University, 
College Station, TX77843, USA}

\address[ad2]{Department of Materials Science \& Engineering, Texas A\&M University, 
College Station, TX77843, USA}

\address[ad3]{Department of Mechanical Engineering, University of Houston, Houston, TX77204, USA}


\begin{abstract}
This work presents a three-dimensional constitutive model for shape memory alloys considering the TRansformation-Induced Plasticity (TRIP) as well as the Two-Way Shape Memory Effect (TWSME) through a large deformation framework. The presented logarithmic strain based model is able to capture the large strains and rotations exhibited by SMAs under general thermomechanical cycling. By using the martensitic volume fraction, transformation strain, internal stress, and TRIP strain tensors as internal state variables, the model is capable to capture the stress-dependent TRIP generation when SMAs are subjected to a multiaxial stress state, as well as the TWSME for thermomechanically trained SMAs under load-free conditions. A detailed implementation procedure of the proposed model is presented through a user-defined material subroutine within a finite element framework allowing for solving different Boundary Value Problems (BVPs). Comprehensive instruction on calibrating the model parameters as well as the derivation of continuum tangent stiffness matrix are also provided. In the end, the simulated cyclic pseudoelastic and actuation responses by the presented model for a wide range of SMA material systems under both uniaxial and multiaxial stress states are compared against experimental results to validate the proposed modeling capabilities.

\end{abstract}

 
\end{frontmatter}

\section{INTRODUCTION} \label{sec:intro} 
Shape Memory Alloys (SMAs) represent an active/smart material with the ability to recover their pre-defined shape via a diffusionless phase transformation between its high-symmetry, high-temperature austenitic phase and low-symmetry, low-temperature martensitic phase. Due to the high output energy density of SMAs, up to 500 MPa actuation stress and 8\% recoverable strain \citep{otsuka1999}, compared to other active materials such as shape memory polymers and piezoelectrics, their current and potential applications in the biomedical, aerospace, automobile and civil engineering fields are expanding rapidly \citep{hartl2007aerospace,song2010applications,peraza2013opt,jani2014review,karakalas2019}. SMAs have been extensively researched as solid-state actuators to enable adaptive and morphing structures. For examples, an SMA-based beam component has been used as a bending actuator to morph the engine outer shell geometry so that a desired aerodynamic conditions can be achieved during the airplane take-off and cruise regime \citep{hartl2007aerospace}. SMA-based torque tubes have been used as rotation actuators to deploy and retract solar panels for small satellites \citep{wheeler2015}, and also used as rotational actuators to rotate the trailing edge wing flap during an airplane on-fly test \citep{mabe2014,calkins2016}.

The majority of engineering applications require SMAs experiencing a large number of loading cycles involving repeated phase transformations, which brings the increasing necessity to understand SMA material response under thermomechanical cycling. Many experimental results \citep{strnadel1995a, strnadel1995b, Bo1999_TRIP_2, lagoudas2004TRIP, wheeler2014} indicate that SMAs exhibit an evolving rather than stable material response under cyclic loading. More specifically, transformation characteristics of SMAs, e.g., the shape of stress/thermal hysteresis, transformation temperatures, transformation strain magnitude, usually shift from one cycle to another, and large irrecoverable strains are usually accumulated. Such irrecoverable strains, often called TRIP, are caused by the microstructure changes as a result of the repeated phase transformation. These microstructure changes, including accumulation of dislocation bands, initiation and growth of micro-voids and micro-cracks, and damage accumulation, then effectively results in an observable macroscopic plastic strain, which occurs at an effective stress level much lower than the conventional plastic yielding point \citep{lagoudas2004TRIP}. In addition, TRIP strain evolves with different rates throughout the entire material fatigue life state. It has been shown in the SMA fatigue test results \citep{wheeler2014characterization} that the TRIP strain grows drastically during the very first hundreds of loading cycles then tends to increase in a stabilized trend until the material reaches the failing point at the end. As the most of engineering applications require actuators functioning in stable material behavior, SMAs are usually subjected to a training process, i.e., repeated thermal/stress cycling, to stabilize their behavior before being used as actuation components. As of now, many modeling efforts have been devoted to predicting the evolving characteristics of SMAs under cyclic thermomechanical loading.

A large number of legacy models have been proposed to predict the stable SMA material response. A thorough review of these works can be found from literature \citep{leclercq1996,boyd1996,patoor1996,tham2001,patoor2006,zaki2007,lagoudas2008,arghavani2011,lagoudas2012,kelly2016,wang2017ijes,xu2017AIAA,xu2018asme}. In general, SMA models considering irrecoverable strains can be categorized into two types. One type of SMA models describe conventional plasticity due to the activation of slip systems at sufficiently high-stress levels in either pure austenite or martensite phase. Modeling efforts fall into this type can be obtained from publication \citep{wang2008,hartl2009,jiang2016,scalet2019}. Another type of models concerns irrecoverable strains as TRIP caused by repeated phase transformations wherein the stress levels are much lower than the material yielding point. It is noted that the focus of this work falls into the second type. Many commonly cited SMA models have been proposed to capture such evolving response feature, and a subset of them are briefly reviewed here. The earliest models dealing with TRIP were presented by \cite{lim1994} and \cite{tanaka1995} to capture the cyclic loading effect on the SMA phase transformation characteristics. Later on, based on the micromechanics averaging method on an SMA representative volume element, \cite{Bo1999_TRIP_1, Bo1999_TRIP_3} proposed a model accounting for the one-dimensional TRIP strain accumulation and the generation of TWSME under actuation cycling. \cite{lexcellent2000} further extended their early SMA model describing stable material response \citep{lexcellent1996} to consider the irrecoverable strains by introducing two additional internal state variables, i.e., the volume fraction of self-accommodated and oriented martensite. Afterward \cite{lagoudas2004TRIP} proposed a model accounting for TRIP as well as the shape and size of the hysteresis during pseudoelastic cycling. Also, \cite{zaki2007} proposed a model considering the TRIP in the case of cyclic pseudoelastic loading by using additional internal state variables, such as internal stress, TRIP strain, and accumulated martensite volume fraction. Other similar and recent modeling efforts can also be found from the works \citep{auricchio2007,saint2009,yu2013,barrera2014,yu2015,xu2017TRIP,xu2019aiaa}.

\begin{figure}[t]	
	\begin{center}
		\includegraphics[width=1.0\columnwidth]{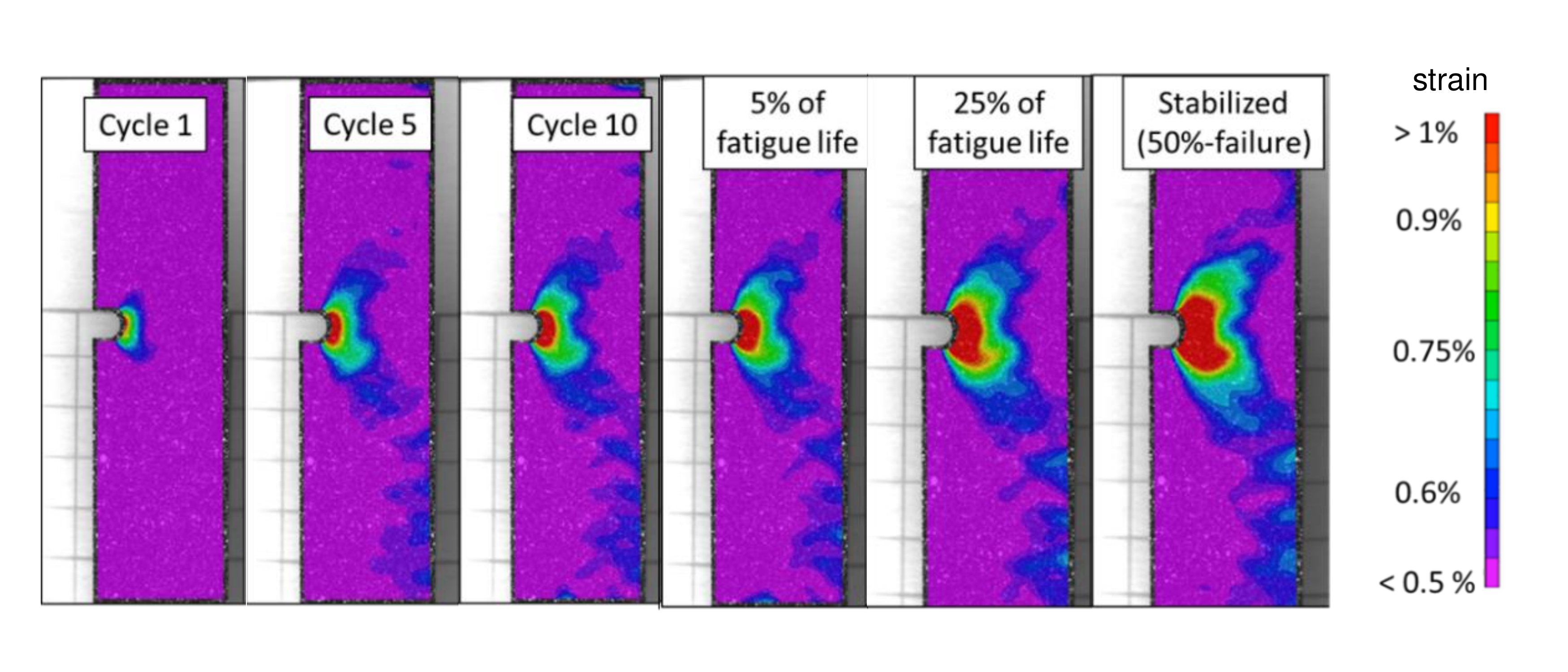}	
	\end{center}
	\vspace*{-0.5cm}
	\caption{Evolution of the summation of transformation and TRIP strain in the loading direction for a notched NiTi plate subjected to thermal cycling under constant load. Referenced from \cite{wheeler2017actuation}.}
	\label{fig:MultiTRIP}
\end{figure}

Although many of the proposed models have enabled researchers to study the evolving material behaviors of SMAs, the majority of them are insufficient in their capacity to consider the following critical features. (\romannumeral 1) The first feature of the currently available models in need of improvement is their small deformation assumption based on infinitesimal strain theory. This assumption may be acceptable for SMA material systems, such as Ni-rich or NiTiHf SMAs, where the summation of all strains, including elastic, transformation, and TRIP strains, is below 3\%. However, it has been reported that nearly 30\% or even higher TRIP strains are observed during the lifetime of near equiatomic NiTi SMA-based actuators \citep{wheeler2014}. Also with the presence of cracks in SMAs, the strain regime in front of the crack tip can easily go beyond 10\% \citep{Haghgouyan2016,Haghgouyan2019}. In the presence of such large strain, a finite strain model is needed to account for the exhibited large strains to provide an accurate structural response of SMA-based multifunctional components. (\romannumeral 2) The second important aspect of many current models in need of improvement is the TWSME characteristic exhibited by trained SMAs at load-free conditions. Because of the required training process to stabilize the response of as-received SMAs before used as actuators, permanent changes such as dislocation bands, accumulation of defects/damage and retained martensite variants are usually introduced into the material's microstructure, which then results in the generation of an internal stress field oriented in the same direction as the applied load. Thereafter the generated internal stress field is capable to induce the oriented phase transformation under pure thermal cycling without applying any external mechanical loads, i.e., the TWSME. Such unique TWSME property of SMAs have tremendous engineering potentials. For instance, it allows for mounting and dismounting of SMA-based connectors and couplers in an easy procedure by just heating and cooling without pre-stressing \citep{niccoli2017,tabesh2018}. (\romannumeral 3) The last major contribution of this work is the consideration of stress-dependent TRIP evolution under multiaxial stress state. The majority of applications require the functionality of actuators under multiaxial stress state originated from geometry curvatures or installment required discontinuities, such as notches and holes. TRIP strain under such multiaxial stress state evolves quite differently compared to that in the uniaxial loading case. Refer to Fig. \ref{fig:MultiTRIP} for the Digital Image Correlation (DIC) strain results of a notched NiTi plate under cyclic actuation loadings, it revealed that the TRIP strain evolved in a much faster rate at the stress concentration region than the less stressed part. Despite the importance of the above mentioned facts, they have rarely been addressed among existing models from available literature through a unified modeling framework. 


In order to address the three aforementioned critical features, a three-dimensional finite strain constitutive model for SMAs considering TRIP and TWSME is proposed in this work. The presented modeling effort is developed based on the legacy SMA model \citep{boyd1996,lagoudas2012,xu_2019sms} and largely inspired by its continuous development considering TRIP \citep{Bo1999_TRIP_3,lagoudas2004TRIP,xu2019aiaa}. The proposed logarithmic strain based model is able to capture the large strains and rotations exhibited by SMAs under general thermomechanical cycling. By using the martensitic volume fraction, transformation strain, internal stress, and TRIP strain tensors as internal state variables, the model is also able to capture the stress-dependent TRIP evolution when SMAs are under a multiaxial stress state, and the TWSME for thermomechanically trained SMAs due to the generation of internal stresses under load-free conditions.

In summary, the paper is organized as follows. In Section~\ref{Preliminary}, kinematic preliminaries used in the model formulation are presented. Section \ref{sec:formulation} focuses on the model development that incorporates the three important features as mentioned earlier. In section \ref{sec:imp}, the detailed implementation procedure for the proposed model is described by using a user-defined material subroutine (UMAT) through the finite element software Abaqus. Thereafter, numerical examples are analyzed to demonstrate the proposed modeling capabilities in Section \ref{Result}. Conclusions are summarized in Section \ref{conc}. In the  \ref{sec:Jacobian}, the explicit derivation of the continuum tangent stiffness matrix is provided, and \ref{sec:calibration} presents a detailed instruction on how to calibrate all the model parameters based on available experimental data.

\section{PRELIMINARY} \label{Preliminary}
\subsection{Kinematics}
Let the position of a material point $\mathcal{P}$ from body $\mathcal{B}$ defined by a vector $ \mathbf{X} $ in the reference (undeformed) configuration at time $ t_{0} $, and vector $ \mathbf{x} $ represents the position of that material point in the current (deformed) configuration at time $ t $. It is well-known that the deformation process of this material point $\mathcal{P}$ from reference to current configurations can be described by using the deformation gradient tensor $\mathbf{F}(\mathbf{x},t)$,
\begin{equation}\label{Deformation}
\mathbf{F}(\mathbf{x},t) =\frac{\partial \mathbf{x}}{ \partial \mathbf{X}}  
\end{equation}
also, the velocity field $\mathbf{v}$ of material point $\mathcal{P}$ can be defined as,
\begin{equation}\label{Velocity}
\mathbf{v} = \dfrac{d \mathbf x}{d t} = \dot{\mathbf x}
\end{equation}
the velocity gradient $\mathbf{L}$ can be derived based on the velocity field $\mathbf{v}$ as,
\begin{equation}\label{eq:V_gradient}
\mathbf{L}= \frac{\partial \mathbf{v}}{ \partial \mathbf{x}}= \mathbf{\dot{F}}\mathbf{F} ^{-1}  
\end{equation}
the polar decomposition for the deformation gradient $\mathbf F$ is also known as follows,
\begin{equation}\label{Polar_Dec}
\mathbf{F  = UR = VR}
\end{equation}
in which $\mathbf{R}$ is the orthogonal rotation tensor, $\mathbf{U}$ and $\mathbf{V}$ are called the right (or Lagrangian) and the left (or Eulerian) stretch tensors respectively. The right Cauchy-Green tensor $\mathbf{C}$ and the left Cauchy-Green tensor $\mathbf{B}$ can be obtained as follows,
\begin{equation}\label{eq:RCG}
\mathbf{C} = \mathbf{F^{T}F} =\mathbf{U}^2 
\end{equation}
\begin{equation}\label{eq:LCG}
\mathbf{B} = \mathbf{FF^{T}} =\mathbf{V}^2 
\end{equation}
the so called logarithmic (or Hencky/true) strain can be expressed in its Lagrangian form $ \mathbf{H} $ and its Eulerian form $ \mathbf{h} $ as,
\begin{equation}\label{eq:L_LogStrain}
\mathbf{H} = \frac{1}{2} \ln(\mathbf{ {C}}) =\mathbf{\ln ({U})}
\end{equation}
\begin{equation}\label{eq:E_LogStrain}
\mathbf{h} = \frac{1}{2} \ln(\mathbf{ {B}}) =\mathbf{\ln ({V})}
\end{equation}

It is also known that the velocity gradient $\mathbf{L}$ can be additively decomposed into a symmetric part, i.e., the rate of deformation tensor $\mathbf{D}$, plus an anti-symmetric part, i.e., the spin tensor $\mathbf{W}$,
\begin{equation}\label{eq:S_P_tensor}
\mathbf{L = D + W}, \qquad
\mathbf{D} =\dfrac{1}{2} \mathbf{(L+L^{T})}, \vspace{5pt}  \qquad 
\mathbf{W} =\dfrac{1}{2} \mathbf{(L-L^{T})}  
\end{equation}
%


\subsection{Logarithmic rate and Logarithmic spin}\label{Logarithmic}
Two kinematic assumptions, i.e., the multiplicative decomposition of deformation gradient $\mathbf{F}$ and the additive decomposition of the rate of deformation tensor $\mathbf{D}$, are usually used in finite deformation theory. Hyperelastic constitutive relation is often used in multiplicative models while hypoelastic constitutive equation is utilized for additive models. For a long time, the rate form hypoelastic constitutive theory has been criticized for its failure to be fully integrable to describe a simple recoverable elastic behavior. Many spurious phenomena, such as shear stress oscillation and residual stress errors are observed in simple elastic deformation using objective rates, this includes many well-known objective rates such as Zaremba-Jaumann rate, Green-Naghdi rate, and Truesdell rate, etc.\citep{xiao2006}. However, such aforementioned issues regarding objective rates are resolved by the logarithmic rate proposed by \cite{xiao1997,xiao1997hypo,xiao2006,bruhns1999self,bruhns2001large,bruhns2001self,meyers2003elastic,meyers2006choice}. As proved in their work, the logarithmic rate of Eulerian logarithmic strain $\mathbf{h}$ is identical with the rate of deformation tensor $\mathbf{D}$, by which a hypoelastic model can be exactly integrated into an finite strain elastic model (\cite{xiao1997}). This unique relationship between the logarithmic strain $\mathbf{h}$ and the rate of deformation tensor $\mathbf{D}$ can be expressed as,
\begin{equation}\label{eq:Log_strain_rate}
\mathring{\mathbf{h}}^{log} = \dot{\mathbf{h}}+\mathbf{h} \mathbf{ \Omega}^{log}-\mathbf{ \Omega}^{log}\mathbf{h}= \mathbf{D}
\end{equation}
where $ \mathbf{ \Omega}^{log} $ is the logarithmic spin introduced by \cite{xiao1997},
\begin{equation}\label{eq:Log_spin}
\mathbf{\Omega}^{log} = \mathbf{W}+ \sum_{i \neq j}^{n}  \big(\frac{1+(\lambda_{i}/\lambda_{j})}{1-(\lambda_{i}/\lambda_{j})}+\frac{2}{\ln (\lambda_{i}/\lambda_{j})}\big) \mathbf{b}_i \mathbf{D} \mathbf{b}_j
\end{equation}
in which $\lambda_{i,j} (i,j=1,2,3) $ are the eigenvalues of left Cauchy-Green tensor $ \mathbf{B} $, $ \mathbf{b}_{i}, \mathbf{b}_{j} $ are the corresponding subordinate eigenprojections. Additionally, the second-order rotation tensor $\mathbf{R}^{log}$ can be obtained by using the logarithmic spin tensor $\mathbf{\Omega}^{log}$ after solving the following tensorial differential equation. The initial condition $\mathbf{R}^{log}|_{t=0}= \mathbf I$ is often assumed.
\begin{equation}\label{eq:R_def}
{\mathbf{\Omega}^{log}}=\dot{\mathbf{R}}^{log}(\mathbf{R}^{log})^T
\end{equation} 

Furthermore, equation (\ref{eq:Log_strain_rate}) can be integrated to the following equation via the corotational integration scheme with the initial condition $\mathbf{h}|_{t=0}= \mathbf 0$, (\cite{khan1995continuum}), 
\begin{equation}\label{eq:h-D}
\mathbf{h} = \int_{\text{corot.}}\mathbf{D} ~\text{d}t =(\mathbf{R}^{log})^T \bigg (\int_{0}^{t} \mathbf{R}^{log}\mathbf{D}^{e} (\mathbf{R}^{log})^T\text{d}t' \bigg) \mathbf{R}^{log}  
\end{equation}
%

\subsection{Additive decomposition of logarithmic strain }\label{AdditiveStrain}
This part addresses the kinematic assumption of additive decomposition of logarithmic strain. First, the rate of deformation tensor $\mathbf D$ is additively decomposed into three parts,\footnote{The thermal strain part is small compared to the other major strains, hene it is omitted here for the sake of simplicity.} i.e., elastic part, transformation part, and TRIP part. From energy point of view, additive decomposition of $\mathbf{D}$ can be interpreted as the  stress power provided from outside is split into an elastic part stored inside the material, a dissipative part associated with phase transformation process, and another dissipative part related to transformation-induced plastic deformation.
\begin{equation}\label{eq:add_D}
\mathbf{D}=\mathbf{D}^{e}+\mathbf{D}^{tr}+\mathbf{D}^{tp}
\end{equation}

Based on equation (\ref{eq:Log_strain_rate}), the elastic part $\mathbf{D}^{e}$, transformation part  $\mathbf{D}^{tr}$  and plastic part $\mathbf{D}^{tp}$ can be rewritten as  $\mathring{\mathbf{h}}^{e\_log}$, $\mathring{\mathbf{h}}^{tr\_log}$ and $\mathring{\mathbf{h}}^{tp\_log}$ respectively,
\begin{equation}\label{eq:add_h_rate1}
\mathring{\mathbf{h}}^{e\_log}=\mathbf{D}^{e};~~\mathring{\mathbf{h}}^{tr\_log}=\mathbf{D}^{tr};~~\mathring{\mathbf{h}}^{tp\_log}=\mathbf{D}^{tp}
\end{equation}
the following equation can be obtained by combining equations (\ref{eq:Log_strain_rate}), (\ref{eq:add_D}) and (\ref{eq:add_h_rate1}),
\begin{equation}\label{eq:add_h_rate2}
\mathring{\mathbf{h}}^{log}=\mathring{\mathbf{h}}^{e\_log}+\mathring{\mathbf{h}}^{tr\_log}+\mathring{\mathbf{h}}^{tp\_log}
\end{equation}
applying corotational integration on equation (\ref{eq:add_h_rate2}) gives,
\begin{subequations}
	\begin{align}
	\mathbf{h}^{e}  &= \int_{\text{corot.}}\mathbf{D}^{e~} ~\text{d}t =(\mathbf{R}^{log})^T \bigg(\int_0^\tau \mathbf{R}^{log}\mathbf{D}^{e} (\mathbf{R}^{log})^T\text{d}\tau \bigg ) \mathbf{R}^{log}  \\
	\mathbf{h}^{tr} &= \int_{\text{corot.}}\mathbf{D}^{tr} ~\text{d}t =(\mathbf{R}^{log})^T \bigg (\int_0^\tau \mathbf{R}^{log}\mathbf{D}^{tr} (\mathbf{R}^{log})^T\text{d}\tau \bigg ) \mathbf{R}^{log}   \\
	\mathbf{h}^{tp} &= \int_{\text{corot.}}\mathbf{D}^{tp} ~\text{d}t =(\mathbf{R}^{log})^T \bigg (\int_0^\tau \mathbf{R}^{log}\mathbf{D}^{tp} (\mathbf{R}^{log})^T\text{d}\tau \bigg ) \mathbf{R}^{log}  	
	\end{align}
	\label{eq:h_coro}
\end{subequations}

Combing equations (\ref{eq:h-D}), (\ref{eq:add_D}), (\ref{eq:add_h_rate2}) and (\ref{eq:h_coro}), the following kinematic equation on total logarithmic strain can be received. Namely, the total logarithmic strain is additively split into an elastic part, a transformation part, as well as a TRIP part
\begin{equation}\label{eq:add_h}
\mathbf{h}=\mathbf{h}^{e}+\mathbf{h}^{tr}+\mathbf{h}^{tp}
\end{equation}

\section{MODEL FORMULATION}\label{sec:formulation}
\subsection{Thermodynamic potential}\label{subsec:Gibbs}
Based on the classical thermodynamic framework for dissipative materials, the development of the proposed model starts with the formulation of an explicit thermodynamic potential. To that end, a quadratic form Gibbs free energy is proposed as a continuous function based on Kirchhoff stress tensor $ \bm{\uptau} $, temperature $ T $, and a set of internal state variables $\mathbf{\Upsilon}=\{\xi,\mathbf{h}^{tr},\mathbf{h}^{tp},\bm{\beta}\}$, in which they are martensitic volume fraction $ \xi $, transformation strain tensor  $ \mathbf{h}^{tr} $, TRIP strain tensor $ \mathbf{h}^{tp} $, and internal stress tensor $\bm{\beta}$. The martensitic volume fraction $ \xi $ ranging $ 0 \leqslant \xi \leqslant 1 $ is used for differentiating the two material phases of SMA. Specifically, $\xi=0$ represents pure austenitic phase while $\xi=1$ indicates pure martensitic phase. The $ \mathbf{h}^{tr} $ accounts for the inelastic yet recoverable strain associated with the phase transformation, $ \mathbf{h}^{tp} $ is used to represent the irrecoverable transformation-induced plastic strain, and $\bm{\beta}$ is used to consider the internal stress field generated inside the material as a result of the training process. The following explicit Gibbs free energy expression for $G$ is given as,
\begin{equation}\label{GIBBS_explicit} 
\begin{aligned}
G=  -\dfrac{1}{2 \rho_{0}} \bm{\uptau} : \mathcal{S}\bm{\uptau} - \dfrac{1}{\rho_{0}}  \bm{\uptau} :[~\bm{\alpha}(T-T_0)+\mathbf{h}^{tr}+\mathbf{h}^{tp}]  -\frac{1}{\rho_{0}} \int^{\xi}_0 (\bm{\beta}:\frac{\partial\mathbf{h}^{tr}}{\partial \tau})d\tau        +c \Big[(T-\\T_0)-T\ln (\dfrac{T}{T_0}) \Big]-s_0 T+u_0+\dfrac{1}{\rho_0}f(\xi)
\end{aligned}
\end{equation}

In which, $\mathcal{S}$ is the fourth-order effective compliance tensor that can be calculated by using the rule of mixture as per equation (\ref{eq:S_mix}), $\mathcal{S}^A$ is the compliance tensor for austenitic phase while $\mathcal{S}^M$ is for martensitic phase, $\Delta\mathcal{S}$ represents the phase difference of the compliance tensor. Additionally, the effective stiffness tensor $\mathcal{C}$ can be obtained by taking the inverse of the above effective compliance tensor, {i.e.}, $\mathcal{C}=\mathcal{S}^{-1}$. $ \bm{\alpha}$ is the second-order thermoelastic expansion tensor, $ c $ is the effective specific heat, $ s_0$ and $ u_0 $ are the effective specific entropy and effective specific internal energy at the reference state. All the aforementioned effective variables are determined by the rule of mixture from equation (\ref{eq:alpha}) to (\ref{eq:u0}). $ T $ represents the temperature at current state, while $ T_0 $ is the temperature at reference state.
\begin{equation}\label{eq:S_mix}
\mathcal{S}(\xi)=\mathcal{S}^A + \xi(\mathcal{S}^M-\mathcal{S}^A)=\mathcal{S}^A + \xi\mathit{\Delta}\mathcal{S}
\end{equation}
\begin{equation}\label{eq:alpha}
\bm{\alpha}(\xi)=\bm{\alpha}^A + \xi(\bm{\alpha}^M-\bm{\alpha}^A)=\bm{\alpha}^A + \xi\Delta\bm{\alpha}
\end{equation}
\begin{equation}\label{eq:c}
{c}(\xi)~= c^A ~+ \xi(c^M-c^A)~~=c^A + \xi\Delta c
\end{equation}
\begin{equation}\label{eq:s0}
{s}_0(\xi)= {s}_0^A + \xi({s}_0^M-{s}_0^A)~~={s}_0^A + \xi\Delta {s}_0
\end{equation}
\begin{equation}\label{eq:u0}
{u}_0(\xi)= {u}_0^A + \xi({u}_0^M-{u}_0^A)~~={u}_0^A + \xi\Delta {u}_0
\end{equation}

A smooth hardening function $f(\xi)$ is included in the Gibbs free energy to account for the polycrystalline hardening effects, such as interactions between different phase variants, imperfections located at the grain boundaries, and the presence of nano-precipitates \citep{lagoudas2008}. Three intermediate material parameters $(a_1,a_2,a_3)$ are introduced in this hardening function, they can be determined as derived in \ref{sec:calibration} by using the known material parameters . Besides, the other four smoothing parameters $(n_1,n_2,n_3,n_4)$ are introduced to effectively treat the smooth transition characteristics at the initiation and completion during phase transformation. The complete form of hardening function is as follows,
\begin{equation}\label{eq:Smooth_hardeing}
\begin{aligned}
f(\xi)=   \begin{cases} \cfrac{1}{2} a_1\Big(  \xi  + \frac{\xi^{n_1+1}} {n_1+1}+ \frac{(1-\xi)^{n_2+1}} {n_2+1} \Big)+a_3\xi ~, \; \dot{\xi}>0, \vspace{5pt} \\ 
\cfrac{1}{2} a_2\Big(  \xi  + \frac{\xi^{n_3+1}} {n_3+1}+ \frac{(1-\xi)^{n_4+1}} {n_4+1} \Big)-a_3\xi ~, \; \dot{\xi}<0 \end{cases}\\
\end{aligned} 
\end{equation}

On basis of the proposed Gibbs free energy, following the classic thermodynamic principles and Coleman-Noll procedure, the constitutive relationships between stress and strain, entropy and temperature can be obtained in equations (\ref{eq:h_Cons_f}) and (\ref{eq:entropy_Cons_f}),
\begin{equation}\label{eq:h_Cons_f}
\mathbf h =  -\rho_{0}\frac{\partial G}{\partial \bm {\uptau}} = \mathcal{S}\bm{\uptau} + \bm\alpha(T-T_0)+ \mathbf{h}^{tr}+ \mathbf{h}^{tp}
\end{equation}   
\begin{equation}\label{eq:entropy_Cons_f}
s =  -\rho_{0}\frac{\partial G}{\partial T} = \frac{1}{\rho_0}  \bm{\uptau} : \bm\alpha  +c \ln (\dfrac{T}{T_0}) -s_0
\end{equation}

The following reduced form of dissipation inequality (\ref{eq:Dissipation_State_V2}) can be derived by substitution of the above constitutive relationships into the Clausius-Planck inequality,
\begin{equation}\label{eq:Dissipation_State_V2}
-\rho_{0}  \frac{\partial G}{\partial \mathbf{h}^{tr}} :\mathring{\mathbf{h}}^{tr} -\rho_{0}  \frac{\partial G}{\partial \mathbf{h}^{tp}} :\mathring{\mathbf{h}}^{tp} -\rho_{0}  \frac{\partial G}{\partial \xi}\dot{\xi} \geqslant 0
\end{equation}
%


\subsection{Evolution law for transformation strain}\label{Evolution_Trans}
This part focuses on proposing the evolution law for transformation strain. Following the maximum dissipation principle, the evolution of transformation strain that tends to dissipate the most energy among all the admissible thermodynamic paths is chosen during the phase transformation process \citep{boyd1996, qidwai2000}. Therefore, it is assumed that the rate change of transformation strain is proportional to the rate change of the martensitic volume fraction $\xi$, and the direction of which is along the deviatoric part of the effective stress tensor. Be noted that in the following evolution law the rate applied on top of the transformation strain is the logarithmic rate in order to account for the principle of objectivity, the 'log' symbol is neglected for the sake of simplicity. Finally, the explicit evolution law for transformation strain is described as follows,
\begin{equation}\label{eq:Trans_Evol}
\begin{aligned}
{\mathring{\mathbf h}}^{tr}= \bm{\Lambda}  \dot{\xi},  \ \  \bm\Lambda=\begin{cases}\bm{\Lambda}_{fwd}, \; \dot{\xi}>0, \vspace{5pt} \\ \bm{\Lambda}_{rev}, \; \dot{\xi}<0, \end{cases}\\
\end{aligned}
\end{equation}
where, $\bm\Lambda_{fwd}$ is the forward transformation direction tensor, and $\bm\Lambda_{rev}$ is the reverse transformation direction tensor. They are defined explicitly as follows,
\begin{equation}\label{eq:Evo_tr}
\bm\Lambda_{fwd}=
\frac{3}{2} H^{cur} 
\frac{\bm{\uptau}^{\text{eff}'}}{\bar{\uptau}^{\text{eff}}}; \ \  \ \     
\Lambda^{rev}=
\frac{\mathbf h^{\textit{tr-r}}}{{\xi}^{\textit{r}}}.
\end{equation}
in the above equations, $ \bm{\uptau}^{\text{eff}} $ is the effective stress tensor defined as equation (\ref{eq:S_eff}) using the summation of Cauchy stress and internal stress. It is worthy to point out that the major difference here compared to available SMA model concerning stable material response is the introduction of this effective stress tensor. It can be seen later on that the TWSME of thermomechanically trained SMAs under load-free conditions can be achieved by using this effective stress term. The evolution law for internal stress tensor $\bm\beta$ is introduced in the later context shortly. 
\begin{equation}\label{eq:S_eff}
\bm{\uptau}^{\text{eff}}=\bm{\uptau}+\bm\beta
\end{equation}
in addition, the deviatoric part of effective stress tensor is defined as { $ \bm{\uptau}^{\text{eff}'} =\bm{\uptau}^{\text{eff}} -{\small \frac{1}{3}}\textrm{tr}(\bm{\uptau}^{\text{eff}})~\mathbf{1} $}, where $\mathbf{1} $ is the second-order identity tensor. The von Mises equivalent effective stress $\bar{\bm\uptau}^{\text{eff}}$ is calculated as follows,
\begin{equation}
\bar{\bm\uptau}^{\text{eff}} ={ \sqrt{{{ \frac{3}{2}}\bm{\uptau}^{\text{eff}'}:\bm{\uptau}^{\text{eff}'}}}} 
\end{equation}

It is common in many of the available models that the magnitude of the recoverable transformation strain is assumed to be the same under different stress levels. Such consideration is valid when the applied stress levels are high enough to generate fully oriented martensitic variants. However, self-accommodated martensitic variants are also generated when the stress levels are not sufficiently high, which renders the value of transformation strain to be a stress-dependent variable. Therefore, the following exponential $H^{\textit{cur}}$ function is introduced to calculate the value of current transformation strain given an effective stress state, where $H^{\textit{max}}$ is the maximum (or saturated) value of transformation strain, $H^{min}$ corresponds to an observable TWSME strain for pre-trained SMAs or some SMAs experiencing particular production process such as extrusion and aging under stress. Besides, $\uptau_{\text{crit}}$ denotes a critical stress value below which $H^{\textit{cur}}=H^{min}$, and $\textit{k}_t$ is a curve-fitting material parameter. The explicit form of $H^{\textit{cur}}$ can be found as follows,   
\begin{equation}\label{eq:Hcur}
H^{\text{cur}}(\bar{\bm\uptau}^{\text{eff}})= \begin{cases} H^{min}+(H^{max}-H^{min})(1-e^{-\text{k}_t ({ \bar{\bm\uptau}^{\text{eff}}}-\uptau_{\text{crit}})}); \hspace{10pt}&\bar{\bm\uptau}^{\text{eff}}> \uptau_{\text{crit}}, \vspace{3pt} \\ 
	H^{min}; &\bar{\bm\uptau}^{\text{eff}} < \uptau_{\text{crit}}, \end{cases}\\
\end{equation}

It is also seen from experimental results \citep{atli2015} that a degradation sometimes exists for the value of maximum transformation strain as a result of the accumulation of retained martensite. In order to extend the model capability to capture this phenomenon, a degradation law is proposed for the maximum transformation strain in equation (\ref{eq:Hmax}), where $H^{max}_i$ and $H^{max}_f$ represent the value of $H^{max}$ before and after the cyclic loading. In addition, $\lambda_1$ is a material parameter governing the degradation trend of $H^{max}$, and $\zeta^d$ called the accumulation of orientated martensitic volume fraction is introduced shortly in the later subsection. 
\begin{equation}\label{eq:Hmax}
H^{max} =  H^{max}_f+(H^{max}_i-H^{max}_f) e^{-\lambda_1\zeta^d}
\end{equation}
 

\subsection{Evolution law for TRIP strain}\label{Evolution_Plastic}
An evolution law for the TRIP strain is proposed in this subsection in order for the model to capture the irrecoverable strain exhibited by SMAs under cyclic thermomechanical loading. Before the detailed formulation is presented, a major assumption is postulated, i.e., among the total martensitic phase transformation, only the oriented transformation portion contributes to the generation of TRIP. This assumption is built upon the observation on the experimental results \citep{lagoudas2008} that no macroscopic irrecoverable deformations are perceived for untrained SMAs under load-free thermal cycling. An early one-dimensional form of TRIP evolution law was suggested in the work of \cite{Bo1999_TRIP_3}, and a three-dimensional form was proposed by \cite{lagoudas2004TRIP}. As it was discussed in the introduction, the generation of TRIP strain is highly stress-dependent under multiaxial stress state, and is totally different compared to that in the uniaxial case. However, none of the above mentioned TRIP evolution laws have well addressed this critical feature, nor the other available models to the authors' best knowledge. Therefore, the following evolution law for TRIP strain considering the effect of stress multiaxiality is suggested, 
\begin{equation}\label{eq:Evol_plastic}
\begin{aligned}
{\mathring{\mathbf h}}^{tp}= \bm{\Lambda}^{tp}  \dot{\xi},  \ \  \bm\Lambda^{tp}=\begin{cases} \bm{\Lambda}^{tp}_{fwd}, \; & \dot{\xi}>0, \vspace{5pt} \\ \bm{\Lambda}^{tp}_{rev}, \; &\dot{\xi}<0, \end{cases}\\
\end{aligned}
\end{equation}

In the above equation,  $\bm\Lambda^p_{fwd}$ is the forward TRIP direction tensor, and $\bm\Lambda^p_{rev}$ is the reverse one. They have explicit definitions in equation (\ref{eq:Dir_plastic}). Note that the rate applied on top of $\mathbf h^{tp}$ is again the logarithmic rate. Because TRIP is strongly driven by the phase transformations, it is proposed that the TRIP strain also evolves along the deviatoric part of the effective stress as the transformation strain described as the following equation,
\begin{equation}\label{eq:Dir_plastic}
\bm\Lambda^{tp}_{fwd}=\frac{3}{2} (\dfrac{H^{cur}}{H^{max}})^2 \dfrac{{ \bm{\uptau}^{'\text{eff}}}} {{{\bar{\bm{\uptau}}}^{\text{eff}}} } \dfrac{C_1^p C_2^p}{1+C_2^p \zeta^d}; \quad \quad 
\bm\Lambda^{tp}_{rev}=-(\dfrac{H^{cur}}{H^{max}})^2 \dfrac{\mathbf h^{tr-r}}{{\xi}^{\textit{r}}} \dfrac{C_1^p C_2^p}{1+C_2^p \zeta^d}
\end{equation}

In the preceding evolution equation, the material parameters $C_1^p$ and $C_2^p$ play the major role in dictating the magnitude and evolving trend for TRIP during cyclic thermo-mechanical loading. The usage of the accumulation of oriented martensitic volume fraction $\zeta^d$ is to be consistent with the previously proposed assumption that only the oriented phase transformation contributes to the TRIP strain generation. For a better illustration on the TRIP evolving trend, the rate form equation (\ref{eq:Dir_plastic}) can be integrated as the following algebraic form using a logarithmic function. 
\begin{equation}\label{eq:Evol_htp_exp}
\mathbf h^{tp}=\frac{3}{2} \dfrac{H^{cur}}{H^{max}}C_1^p ~\dfrac{{ \bm{\uptau}^{'\text{eff}}}} {{{\bar{\bm{\uptau}}}^{\text{eff}}} } {\ln} \big(1+C_2^p \zeta^d \big) 
\end{equation} 

As it can be seen from the above equation, the stress-dependent feature for TRIP strain generation is incorporated by multiplying model parameter $C_1^p$ with the ratio (${H^{cur}}$/${H^{max}}$). As it is shown in the latter results section, such consideration together with the usage of $\zeta^d$ in the evolution law enable the model's capability to capture the highly stress-dependent TRIP generation under multiaxial stress condition. It is also worth to point out that the proposed logarithmic function based evolution law indicates there is no saturation point on TRIP, which is closely aligned with the experimental results from actuation cycling \citep{wheeler2017actuation}. While in the previous work \citep{lagoudas2004TRIP}, an exponential function based evolution law was adopted to be aligned with pseudoelastic cycling experimental results, which indicates TRIP is expected to reach a saturation value, and eventually a stable pseudoelastic response is anticipated to be attained after a certain number of mechanically cycling. Lastly, the accumulation of oriented martensitic volume fraction $\zeta^d$ used in this equation is defined as follows,
\begin{equation}\label{eq:zetad}
\zeta^d = \int_0^t |\dot{\xi}^d(t)|dt
\end{equation}
in which the oriented martensitic volume fraction $\xi^d$ is calculated as,
\begin{equation}\label{eq:xid}
\xi^d = \dfrac{H^{cur}}{H^{max}}~\xi
\end{equation}
the relation between the accumulation of oriented martensitic volume fraction $\zeta^d$ and the accumulation of total martensitic volume fraction $\zeta$ is,
\begin{equation}\label{eq:zeta}
\zeta^d = \dfrac{H^{cur}}{H^{max}}~\zeta
\end{equation}

\subsection{Evolution law for internal stress}\label{sub:Evolution_Internal}
In order to meet the proposed goal of capturing the TWSME at load-free conditions for thermomechanically trained SMAs, a second-order internal stress tensor is introduced in this model.  As discussed to some extents in the introduction section, during cyclic thermomechanical loading SMAs experience microstructure changes, such as pileup of dislocation bands driven by phase transformations, initiation and growth of micro-voids and micro-cracks, and damage accumulation, at a stress level below the yielding point. These microstructure changes gradually introduce a local stress field inside the SMA materials, which is described by the proposed internal stress tensor via an effective manner. It is reasonably assumed the internal stress may never go beyond the material yielding point, thus its magnitude is saturated at a maximum point after a certain number of loading cycles. Besides, the evolution direction of the internal stress is determined as the same direction of the applied stress field due to the fact that the induced microstructure changes are caused by the external mechanical loads. On the basis of the above discussions, the following exponential evolution law is proposed for the internal stress as, 
\begin{equation}\label{eq:internal_stress}
\bm{\beta} =  {\sigma}_b~ \dfrac{H^{cur}}{H^{max}} \dfrac{\bm \uptau^{\text{eff}}}{\bar{\bm \uptau }^{\text{eff}}} (1-e^{-\lambda_1\zeta^d}) 
\end{equation}
in which, $\sigma_b$ is a model parameter representing the maximum (or saturated) magnitude of the internal stress. Similar as the TRIP evolution law, the ratio (${H^{cur}}$/${H^{max}}$) and the term $\zeta^d$ are included here to consider the stress-dependent effect on the internal stress evolution, and $\lambda_1$ is a material parameter controlling the evolution trend.


\subsection{Transformation function}\label{Trans_Func}
In this subsection, a transformation function and an associated transformation criterion are defined, upon which the initiation and completion of phase transformation can be determined. 
By substitution of the evolution law for transformation strain (\ref{eq:Evo_tr}) and TRIP strain (\ref{eq:Evol_plastic}) into the reduced form of dissipation inequality (\ref{eq:Dissipation_State_V2}), the following equation is obtained,
%
%
\begin{equation}\label{Dissipation_xi}
\begin{aligned}
\big(\bm\uptau:\bm\Lambda  + \bm\beta:\bm\Lambda + \bm\uptau:\bm\Lambda^{tp}- 
\rho_{0}\frac{\partial G}{\partial \xi}\big)\dot{\xi}=\pi\dot{\xi}\geqslant 0 
\end{aligned}
\end{equation}
in which the quantity $\pi$ is called the general thermodynamic driving force conjugated to $\xi$ in the proposed model. The product by substitution of Gibbs free energy (\ref{GIBBS_explicit}) into the above equation (\ref{Dissipation_xi}) yields the explicit expression for $ \pi $ in the following equation, where $\mathit{\Delta}\mathcal{S}, \Delta \bm{\alpha}, \Delta c, \Delta s_0, \Delta u_0 $ each represents the phase difference between martenstie and austenite for that variable.
\begin{equation}\label{eq:Driving_Force}
\begin{aligned}
\pi=(\bm\uptau+ \bm\beta):\bm\Lambda + \bm\uptau:\bm\Lambda^{tp}+
\dfrac{1}{2}\bm\uptau:{\Delta}\mathcal{S}\bm\uptau+\bm\uptau:{\Delta}\bm{\alpha}(T-T_0)+ \rho\Delta s_0 T  -\rho\Delta c \big[ T \\ -T_0-T\ln(\dfrac{T}{T_0}) \big ] - \rho\Delta u_0 - \frac{\partial f}{\partial \xi}
\end{aligned}
\end{equation}

Following the development in the early models \citep{boyd1996,lagoudas2012}, it is assumed that the forward (reverse) phase transformation initiates whenever the thermodynamic driving force $ \pi $ ($-\pi $) reaches a critical value $Y $ ($-Y$), and $\pi$ ($-\pi$) is always below such critical value as long as the forward (reverse) phase transformation is not completed, upon which the following transformation function $\Phi$ is defined, 
\begin{equation}\label{eq:Transfor_Fun}
\normalfont{\Phi}=\begin{cases}~~\pi - Y, \; \dot{\xi}>0, \vspace{5pt} \\ -\pi - Y, \; \dot{\xi}<0, \end{cases}\\
\end{equation}

A further improvement in the critical value $Y$ is suggested from \cite{lagoudas2012}, see equation (\ref{eq:Critical_Y}), wherein $Y$ is constructed to be stress-dependent quantity with a reference critical value $Y_0$ and an additional parameter $D$. Such consideration enables the model to capture the different size of hysteresis loops when SMA materials are subjected to various applied stress levels. This capability is provided through capturing the different slopes $C_A, C_M$ in the effective stress-temperature phase diagram.
\begin{equation}\label{eq:Critical_Y}
Y(\bm{\sigma}) = \begin{cases}Y_0 + D\bm\sigma:\bm\Lambda_{\textit{fwd}}, \; \dot{\xi}>0, \vspace{5pt} \\ Y_0 + D\bm\sigma:\bm\Lambda_{\textit{rev}}, \; \dot{\xi}<0, \end{cases}\\
\end{equation}
%


In order for the defined transformation functions and the corresponding transformation criteria to satisfy the principle of maximum dissipation, the following Kuhn-Tucker constraint conditions have to be met,
\begin{equation}\label{eq:Kuhn-Tucker}
\begin{aligned}
\dot{\xi} \geqslant 0; \quad \Phi(\bm\uptau,T,\xi)= ~~\pi - Y \leqslant 0;  \quad  \Phi\dot{\xi}=0;\\
\dot{\xi} \leqslant 0; \quad \Phi(\bm\uptau,T,\xi)= -\pi - Y \leqslant 0; \quad   \Phi\dot{\xi}=0;
\end{aligned}
\end{equation}
%

\section{IMPLEMENTATION}\label{sec:imp}
\begin{figure}[H]
	\centering
	\includegraphics[width=1.0\textwidth]{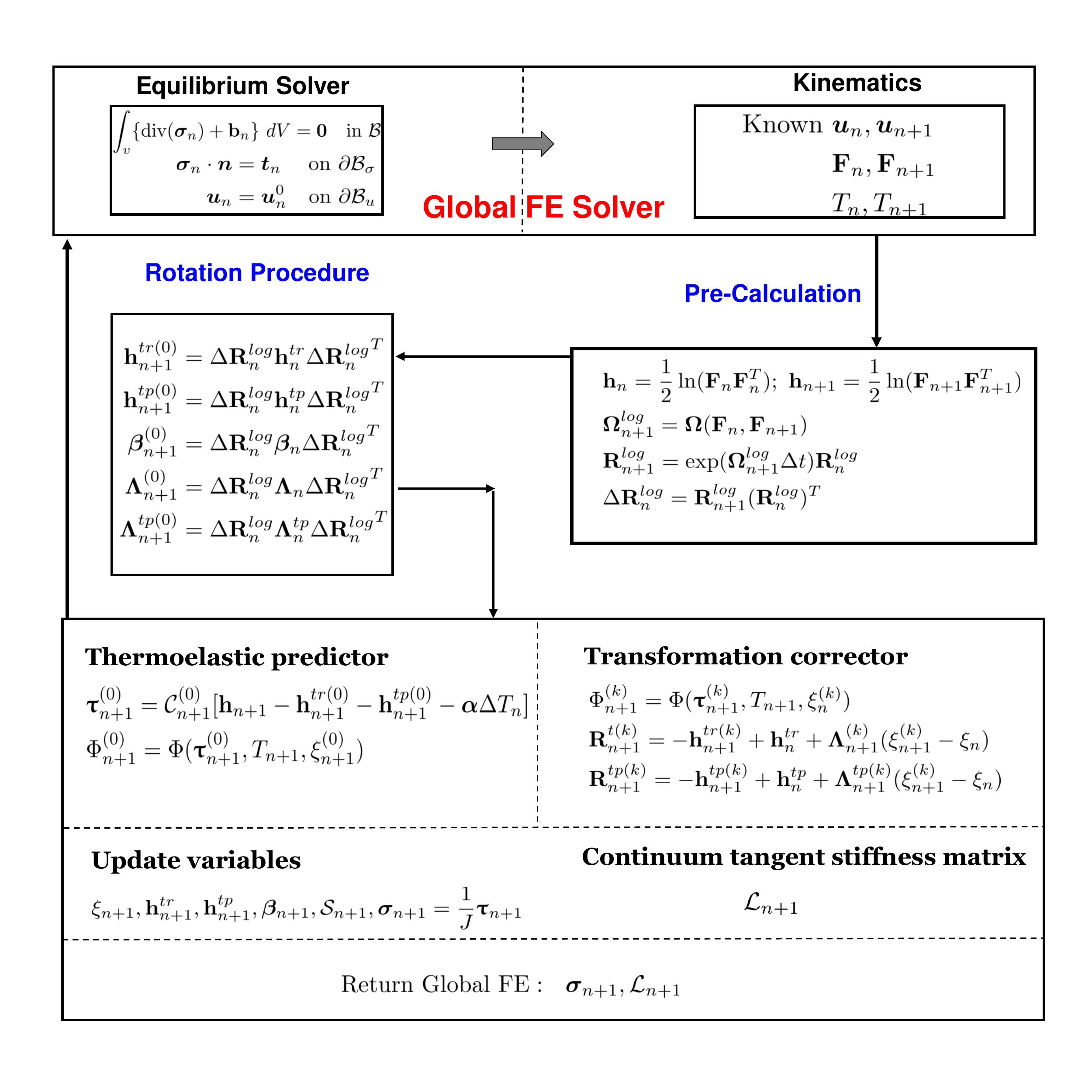}\vspace{-0.5cm}
	\caption{UMAT flowchart for all the variables used in the model during the implementation within the finite element framework.}
	\label{fig:Flowchart}
\end{figure}

This section discusses the implementation of the above proposed model into a numerical environment through a user-defined material subroutine (UMAT), the adopted finite element tool Abaqus equipped with UMAT is then used for solving different BVPs. As shown in Fig. \ref{fig:Flowchart}, all the UMAT variables used in the this model are presented via a flowchart. This implementation procedure follows the knowledge of Return Mapping Algorithm (RMA) presented in the available publications \citep{qidwai2000IMP, lagoudas2008}. The general goal of the UMAT is, given an increment of strain and temperature from the finite element solver, to calculate the output stress values by using the constitutive relationships (\ref{eq:h_Cons_f}) and (\ref{eq:entropy_Cons_f}) with all the internal state variables conforming to the Kuhn-Tucker consistency constraints (\ref{eq:Kuhn-Tucker}). The implementation in general consists of two major procedures, one is called the Thermoelastic-predictor, and the other is called the Transformation-corrector. It is worthy to point out that the input variables used for this implementation collected from the global finite element solver are the temperatures ($T_n$,$~T_{n+1}$), and deformation gradients at the current and next step ($\mathbf{F}_n$, $\mathbf{F}_{n+1}$). As it was discussed in the work of \cite{xu_2019sms}, such non-trivial considerations on the kinematics allow for the implementation of the model to get rid of the accumulated stress errors as a result of using other non-integrable objective rates. The effects of such accumulated stress errors on the cyclic thermomechanical response of SMAs are analyzed in detail from \cite{xu_2019sms}.

\newpage
\captionof{table}{The summary of the UMAT implementation.}\vspace{-0.1cm}
\begin{tabular}{lp{0.9\textwidth}}\toprule[0.3mm]
	& 1.\textit{Initialization}
	\vspace{-0.5cm}
	\begin{itemize}
		\itemsep-0.6em 
		\item Conduct pre-calculation and rotation procedures.
		\item $k=0;  \xi^{(0)}_{n+1}=\xi_{n};  \mathbf{h}^{tr(0)}_{n+1}=\mathbf{h}^{tr}_{n}; \mathbf{h}^{tp(0)}_{n+1}=\mathbf{h}^{tp}_{n}; \bm\beta^{(0)}_{n+1}=\bm\beta_{n} $
	\end{itemize}\vspace{-1.2cm}\\	
	& 2.\textit{Thermoelastic Predictor}
	\vspace{-0.5cm}
	\begin{itemize}
		\itemsep-0.6em 
		\item $\bm\uptau^{(0)}_{n+1}= \mathcal{C}^{(0)}_{n+1}[  \mathbf h_{n+1}  - \mathbf{h}^{tr(0)}_{n+1}  - \mathbf{h}^{tp(0)}_{n+1} -\bm\alpha(T_{n+1}-T_{0}) ] $
		\item Calculate $\Phi^{(k)}_{n+1}$	.
		\item IF $\Phi^{(0)}_{n+1} \leqslant tol  $, GOTO 4 (thermoelastic response).
		\item IF $\Phi^{(0)}_{n+1} > tol  $, GOTO 3 (transformation happens).
	\end{itemize}\vspace{-1.2cm}\\
	%
	& 3.\textit{Transformation Corrector}
	\vspace{-0.5cm}
	\begin{itemize}
		\itemsep-0.6em 
		\item Calculate residual matrix \newline
		$\mathbf R_{n+1}^{tr(k)} = -{\mathbf{h}}^{tr(k)}_{n+1}+{\mathbf{h}}^{tr}_{n}+\bm\Lambda_{n+1}^{(k)}(\xi_{n+1}^{(k)}-\xi_{n})$\newline
		$\mathbf R_{n+1}^{tp(k)} =-{\mathbf{h}}^{tp(k)}_{n+1}+{\mathbf{h}}^{tp}_{n}+\bm\Lambda_{n+1}^{tp(k)}(\xi_{n+1}^{(k)}-\xi_{n})$\newline
		$\Phi^{(k)}_{n+1} ~~= \Phi(\bm{\uptau}^{(k)}_{n+1},T_{n+1},\xi^{(k)}_{n+1})$
		\item Perform Newton-Raphson iterations in equation (\ref{eq:Jacob_matrix}) to obtain
		$\Delta \xi_{n+1}^{(k)}, {\Delta\mathbf{h}}^{tr(k)}_{n+1}, {\Delta\mathbf{h}}^{tp(k)}_{n+1}$.
		\item Update variables $\xi_{n+1}^{(k+1)},{\mathbf{h}}^{tr(k+1)}_{n+1},{\mathbf{h}}^{tp(k+1)}_{n+1},\mathcal{S}^{(k+1)}_{n+1}$\newline
		$\xi_{n+1}^{(k+1)}       =\xi_{n+1}^{(k)}+\Delta\xi_{n+1}^{(k)}$\newline
		${\mathbf{h}}^{tr(k+1)}_{n+1}={\mathbf{h}}^{tr(k)}_{n+1}+\Delta{\mathbf{h}}^{tr(k)}_{n+1} $\newline
		${\mathbf{h}}^{tp(k+1)}_{n+1}={\mathbf{h}}^{tp(k)}_{n+1}+\Delta{\mathbf{h}}^{tp(k)}_{n+1}$\newline
		$\mathcal{S}^{(k+1)}_{n+1}    =\mathcal{S}^{A}+\xi_{n+1}^{(k+1)}\Delta\mathcal S$	
		\item IF $\Phi^{(k+1)}_{n+1} \geqslant tol  $, GOTO step 3, next local iteration $k=k+1$.\newline
		ELSE GOTO step 4, EXIT
	\end{itemize}\vspace{-1.2cm}\\
	
	& 4.\textit{Calculate continuum tangent stiffness $\mathcal{L}$ 
	}.
	\vspace{-0.1cm}
	\begin{itemize}
		\itemsep-0.6em 
		\item ${\mathcal{L}}=\mathcal{C}+\dfrac{[\mathcal{C}({\Delta\mathcal{S}}\bm{\uptau}+ \bm\Lambda +\mathbf\Lambda^{tp})] \otimes  [\mathcal{C} \partial_{\bm\uptau}\Phi]}{\partial_{\xi}\Phi- \partial_{\bm\uptau}\Phi: \mathcal{C}(\Delta\mathcal{S}\bm\uptau+\bm\Lambda+\mathbf\Lambda^{tp})}$
	\end{itemize}\vspace{-0.9cm}\\
	
	& 5.\textit{Update} $\bm{\sigma}_{n+1}$, $\zeta^d_{n+1}$ and $\bm{\beta}_{n+1}$
	\vspace{-0.4cm}
	\begin{itemize}
		\itemsep-0.35em 
		\item $\bm{\sigma}_{n+1}={\small \frac{1}{J}}\bm{\uptau}_{n+1}$, wherein $\bm{\uptau}_{n+1}$ is calculated based on equation (\ref{eq:h_Cons_f}) 
		\item $\zeta^d_{n+1}=\zeta^d_{n}+ \frac{H^{cur}_{n+1}}{H^{max}}|\xi_{n+1}-\xi_{n}|$
		\item $\bm{\beta}_{n+1}={\sigma}_b~  \frac{\bm \uptau^{\text{eff}}_{n+1}}{\bar{\bm \uptau }^{\text{eff}}_{n+1}} (1-e^{-\lambda_1\zeta^d_{n+1}}) $
	\end{itemize}\vspace{-1.1cm}\\
	
	& 6.Exit UMAT and proceed to the global FE solver for the next increment\\	
	\bottomrule[0.3mm]
\end{tabular}

\setlength{\parindent}{2em}

\vspace{0.8cm}
A pre-calculation and a rotation procedure are employed before calling the thermoelastic prediction and transformation correction steps. In the pre-calculation procedure, the total strain at current and next step ($\mathbf{h}_n$, $\mathbf{h}_{n+1}$) are calculated based on ($\mathbf{F}_n$, $\mathbf{F}_{n+1}$) using equation (\ref{eq:E_LogStrain}). The incremental rotation tensor $\Delta\mathbf{R}_n^{log}$ based on the logarithmic rate can be calculated by using the exponential map scheme described in \cite{simo2006}. In the rotation procedure, the tensorial variables stored as solution-dependent quantities including $\mathbf{h}_n$, $\mathbf{h}^{tr}_n$, $\mathbf{h}^{tp}_n$ , $\bm{\beta}_{n}$, $\mathbf{\Lambda}_n$, and $\mathbf{\Lambda}^{tp}_n$ are rotated from the previous $n^{th}$ configuration to the current $(n+1)^{th}$ configuration using $\Delta\mathbf{R}_n^{log}$, thus, to preserve the so-called principle of objectivity. To proceed with the thermoelastic-predictor step, the internal state variables $\mathbf{\Upsilon}^{(0)}_{n+1}=\{\mathbf{h}^{tr(0)}_{n+1},\mathbf{h}^{tp(0)}_{n+1},\bm{\beta}^{(0)}_{n+1},\xi^{(0)}_{n+1}\}$\footnote{$(\cdot)^{(k)}$ represent the local value of that variable at the $k^{th}$ iteration in the transformation correction procedure, here $k=0$ means that this is an  initial guess value in thermoelastic prediction procedure.} at $(n+1)^{th}$ step are assumed to be $\mathbf{\Upsilon}_{n}$ for the initial thermoelastic evaluation. In the case that the Kuhn-Tucker consistency condition is violated, {i.e.}, $\Phi^{(0)}_{n+1} \geqslant 0$, the transformation correction procedure is initiated to attain updated internal state variables to regain consistency. Otherwise, the current $(n+1)^{th}$ step is detected as a thermoelastic response with $\mathbf{\Upsilon}_{n+1}=\mathbf{\Upsilon}^{(0)}_{n+1}$, and the UMAT skips the transformation correction step and continues the rest procedures. A more detailed description of implementation steps is presented in the following context.
\subsection{Thermoelastic Predictor}
This part is a detailed description of the thermoelastic prediction step. Take the $(n+1)^{th}$ step for example, the total strain $\mathbf h_{n+1}$ and temperature $T_{n+1}$ are obtained from the Pre-calculation procedure, and the initial internal state variables $\mathbf{\Upsilon}_{n+1}^{(0)}$ are assumed the same as $\mathbf{\Upsilon}_{n}$ for the initial consistency evaluation, i.e.,
\begin{equation}\label{eq:implement_SDV}
\mathbf{h}^{tr(0)}_{n+1}=\mathbf{h}^{tr}_{n};~~~\mathbf{h}^{tp(0)}_{n+1}=\mathbf{h}^{tp}_{n};~~~\bm\beta^{(0)}_{n+1}=\bm\beta_{n};~~~\xi^{(0)}_{n+1}=\xi_{n}
\end{equation}
on the basis of the above information, the guessed stress value $\bm\uptau^{(0)}_{n+1}$ is calculated through the constitutive equation (\ref{eq:h_Cons_f}),
\begin{equation}\label{eq:implement_stress}
\begin{aligned}
\bm\uptau^{(0)}_{n+1}= \mathcal{C}^{(0)}_{n+1}\Big[  \mathbf h_{n+1}  - \mathbf{h}^{tr(0)}_{n+1}  - \mathbf{h}^{tp(0)}_{n+1}- \bm\alpha^{(0)}_{n+1}(T_{n+1}-T_{0})    \Big]  
\end{aligned}
\end{equation}
the initial value of transformation function $\Phi^{(0)}_{n+1}$ in thermoelastic procedure can be evaluated based on equations (\ref{eq:Driving_Force}) and (\ref{eq:Transfor_Fun}) as follow, 
\begin{equation}\label{eq:implement_phi}
\begin{aligned}
\Phi^{(0)}_{n+1}= \Phi(\bm\uptau^{(0)}_{n+1},T_{n+1},\mathbf{\Upsilon}^{(0)}_{n+1}) 
\end{aligned}
\end{equation}
If the transformation consistency constraints are preserved, i.e., the initial value of transformation function $\Phi^{(0)}_{n+1} \leqslant 0$ \footnote{Usually a small value 'tol' is used for $\Phi^{(0)}_{n+1}\leqslant\text{'tol'}$ evaluation, 'tol' is acceptable to be $10^{-6}$ or a even smaller value.}, no phase transformation is initiated inside SMAs under the current material state, and the current step is detected as a thermoelastic step. If the transformation criterion is violated, i.e. $\Phi^{(0)}_{n+1} \geqslant 0$, the transformation correction procedure is activated in order to restore the consistency constraints (\ref{eq:Kuhn-Tucker}) via retaining updated internal state variables $\mathbf{\Upsilon}^{(k)}_{n+1}$.   
\subsection{Transformation Corrector}
This part focuses on the iterative transformation correction procedure to find a set of updated internal state variables to regain the transformation consistency conditions, i.e., $\Phi^{(k)}_{n+1} \leqslant 0$. Take the $k^{th}$ iteration as an example, the objective is to solve a system of nonlinear equations summarized in equations (\ref{eq:R3}), (\ref{eq:R1}) and (\ref{eq:R2}), where the residuals in the rate form evolution equations (\ref{eq:Trans_Evol}) and (\ref{eq:Evol_plastic}) for transformation strain and TRIP strain can be reformulated into the discretized form as equations (\ref{eq:R1}) and (\ref{eq:R2}),
\begin{equation}\label{eq:R3}
\Phi^{(k)}_{n+1} ~= ~ \Phi(\bm{\uptau}^{(k)}_{n+1},T_{n+1},\xi^{(k)}_{n+1})
\end{equation}%
\begin{equation}\label{eq:R1}
\mathbf R_{n+1}^{tr(k)} = -{\mathbf{h}}^{tr(k)}_{n+1}+{\mathbf{h}}^{tr}_{n}+\bm\Lambda_{n+1}^{(k)}(\xi_{n+1}^{(k)}-\xi_{n})
\end{equation}
\begin{equation}\label{eq:R2}
	\mathbf R_{n+1}^{tp(k)} =-{\mathbf{h}}^{tp(k)}_{n+1}+{\mathbf{h}}^{tp}_{n}+\bm\Lambda_{n+1}^{tp(k)}(\xi_{n+1}^{(k)}-\xi_{n})
\end{equation}

The goal to regain the consistency condition then becomes to satisfy the following convergence inequalities, in which 'tol' means a small convergence value usually can be set as $10^{-6}$.
\begin{equation}\label{eq:NR_Criterion}
\begin{aligned}
|\Phi^{(k)}_{n+1}|\leqslant \text{tol}~; \quad 
|\mathbf R_{n+1}^{tr(k)}|\leqslant\text{tol}~; \quad 
|\mathbf R_{n+1}^{tp(k)}|\leqslant\text{tol}
\end{aligned}
\end{equation}

A standard Newton-Raphson iteration procedure can be utilized to solve the above nonlinear system of  equations. The following equation is usually the explicit form to be iterated wherein the first matrix term in the right-hand side is the inverse of the so-called Jacobian matrix for the nonlinear system of equations.
\begin{equation}\label{eq:Jacob_matrix}
\arraycolsep=1.0pt\def\arraystretch{2}
\left[\begin{array}{c} \Delta \xi_{n+1}^{(k)} \\ {\Delta\mathbf{h}}^{tr(k)}_{n+1} \\ {\Delta\mathbf{h}}^{tp(k)}_{n+1}\end{array} \right] 
=~-\left[\begin{array}{c} 
\dfrac{\partial \Phi^{(k)}_{n+1}}{\partial \xi} ~~ \dfrac{\partial \Phi^{(k)}_{n+1}}{\partial \mathbf{h}^{tr} }~~ \dfrac{\partial \Phi^{(k)}_{n+1}} {\partial \mathbf{h}^{tp} } \\ 
\dfrac{\partial \mathbf R_{n+1}^{tr(k)} }{\partial \xi} ~~ \dfrac{\partial \mathbf R_{n+1}^{tr(k)} }{\partial \mathbf{h}^{tr} }~~ \dfrac{\partial \mathbf R_{n+1}^{tr(k)} } {\partial \mathbf{h}^{tp} } \\ 
\dfrac{\partial \mathbf R_{n+1}^{tp(k)} }{\partial \xi} ~~ \dfrac{\partial \mathbf R_{n+1}^{tp(k)} }{\partial \mathbf{h}^{tr} }~~ \dfrac{\partial \mathbf R_{n+1}^{tp(k)} } {\partial \mathbf{h}^{tp} } \\ 
\end{array} \right]^{-1} 
\left[\begin{array}{c} \Phi^{(k)}_{n+1} \\ \mathbf R_{n+1}^{tr(k)} \\ \mathbf R_{n+1}^{tp(k)} \end{array} \right] 
\end{equation}

During each $k^{th}$ iteration of the Newton-Raphson precedure, the following updated values for the next ${k+1}^{th}$ iteration are obtained for the internal state variables,
\begin{equation}\label{eq:Jacob_matrix_rearrange}
\arraycolsep=1.0pt\def\arraystretch{2}
\left[\begin{array}{c} \xi_{n+1}^{(k+1)} \\ {\mathbf{h}}^{tr(k+1)}_{n+1} \\ {\mathbf{h}}^{tp(k+1)}_{n+1} \end{array} \right] 
=\left[\begin{array}{c} \xi_{n+1}^{(k)} \\ {\mathbf{h}}^{tr(k)}_{n+1} \\ {\mathbf{h}}^{tp(k)}_{n+1} \end{array} \right] 
+\left[\begin{array}{c} \Delta \xi_{n+1}^{(k)} \\ {\Delta\mathbf{h}}^{tr(k)}_{n+1} \\ {\Delta\mathbf{h}}^{tp(k)}_{n+1}\end{array} \right] 
\end{equation}

The Newton-Raphson procedure iterates a maximum number of iterations until the convergence criterion equation (\ref{eq:NR_Criterion}) is satisfied. Once the converged values $\{\mathbf{h}^{tr(k+1)}_{n+1},\mathbf{h}^{tp(k+1)}_{n+1} ,\xi^{(k+1)}_{n+1}\}$ are accepted as the final value for the current material state at the $k^{th}$ iteration, the current transformation correction step is completed.  

\section{RESULTS AND DISCUSSIONS}\label{Result}
In this section, four BVPs are analyzed to test the proposed modeling capabilities of this model considering TRIP and TWSME. These BVPs include both pseudoelastic and actuation cycling for a wide range of SMA material systems, incorporating Ni$_{40}$Ti$_{50}$Cu$_{10}$, Ni$_{49.9}$Ti$_{50.1}$, Ni$_{55}$Ti$_{45}$, and Ni$_{50.3}$Ti$_{29.7}$Hf$_{20}$ with their compositions all in atomic percentage (at.\%). Because the phase transformation characteristics, such as the transformaiton temperatures and strains, shape and size of hysteresis, TRIP strain accumulations, are very various for different material compositions, the fidelity of the proposed model is therefore tested over a variety of SMAs. Three out of the four BVPs are in uniaxial loading conditions while the last one is a multiaxial one.

The first BVP is a Ni$_{40}$Ti$_{50}$Cu$_{10}$ (at.\%) SMA wire under uniaxial cyclic pseudoelastic loading, in which the accumulation of TRIP strain and the stress levels required to initiate the phase transformation after loading cycles are analyzed. Secondly, two BVPs investigate a Ni$_{49.9}$Ti$_{50.1}$ (at.\%) and a Ni$_{50.3}$Ti$_{29.7}$Hf$_{20}$ (at\%) SMA dogbone specimens under uniaxial cyclic actuation loading. Apart from the accumulated TRIP strain, the load-free TWSME responses after the training procedure are also analyzed. Finally, a Ni$_{55}$Ti$_{45}$ (at.\%) plate actuator with a centric hole is chosen to test the model's capability to capture the stress-dependent TRIP evolution under multiaxial stress state. It should be noted that the logarithmic (or ture) strain is adopted in both the simulation and experimental results, and the calibration for the model parameters are based on the true stress-strain curves instead of the engineering one. Under large strain conditions, using an engineering scale stress-strain curve to calibrate the model is expected to cause some discrepancies on material parameters such as Young's modulus and transformation strain as discussed in \cite{xu_2019sms}. Lastly, the simulation of these BVPs are obtained through the commercial finite element software Abaqus, into which the constitutive response of SMAs described above is implemented as a UMAT discussed in the previous section.
%


\begin{table}[t] 
	\centering
	\caption{Model parameters used for NiTiCu SMA under uniaxial cyclic pseudoelastic loading}\label{tab:MP_NiTi_Mechanical}
	\renewcommand{\arraystretch}{0.7}
	\begin{tabular}{c|lr|ll} \toprule
		Type                         &Parameter                        & Value                                   &Parameter            & Value  \\                                       \midrule
		&$E_A$                       & 70   [GPa]                   & $C_A$                                   & 3.4  [MPa/K]\\
		&$E_M$                       & 50   [GPa]                   & $C_M$                                   & 3.4  [MPa/K]\\
		Key material parameters       &$\nu_A=\nu_M$                   & 0.33~~~~~~~~                              & $M_s$                & 264  [K]\\
		12                            &$\alpha_A=\alpha_M$             & 2.2$\times$10$^{-5}$ [K$^{-1}$]                           & $M_f$                &160  [K]\\
		& $ H^\text{max}_i$              & 5\%                            & $A_s$                 &217  [K]\\
		& $ H^\text{max}_f$              & 5\%                            & $A_f $                &290  [K] \\
		& $ H^\text{min}$           & 0\%                         & $k_t$                 &  0.00752 [1/MPa]\\   
		& $\uptau_{\text{crit}}$      &&&                             \\   \midrule
		
		Smooth hardening parameters 		&  $n_1$         &   0.2                &  $n_3$             & 0.4 \\
		4							   &  $n_2$         &   0.3                &  $n_4$        	 & 0.5 \\  \midrule
		
		&${\sigma}_b$                        & 100    [MPa]                      & $\lambda_1$                & 3.5 \\
		TRIP parameters               &$C_1^p$                      & 2.1$\times$10$^{-2}$                     &                  &   \\
		4                     &$C_2^p$                      & 0.17                                 &                 &  \\                              
		\bottomrule
	\end{tabular}
\end{table} 

\subsection{Uniaxial cyclic pseudoelastic loading case}\label{uniaxial_pseudo}
The first BVP describes a Ni$_{40}$Ti$_{50}$Cu$_{10}$ (at.\%) SMA wire under uniaxial pseudoelastic cyclic loading. A bias mechanical load is applied in the longitudinal wire direction from 0 to 550 MPa and then unloaded, the temperature is kept constant at $360$ K throughout the entire procedure. Such loading and unloading process is repeated for 50 cycles. The NiTiCu SMA wire experienced a stress-induced forward phase transformation during the loading regime followed by a reverse phase transformation at the unloading. The material parameters used in this simulation are listed in table \ref{tab:MP_NiTi_Mechanical} based on the calibration from experimental data.

\begin{figure}[h]	
	\begin{center}
		\advance\leftskip-0.5cm
		\includegraphics[width=0.85\columnwidth]{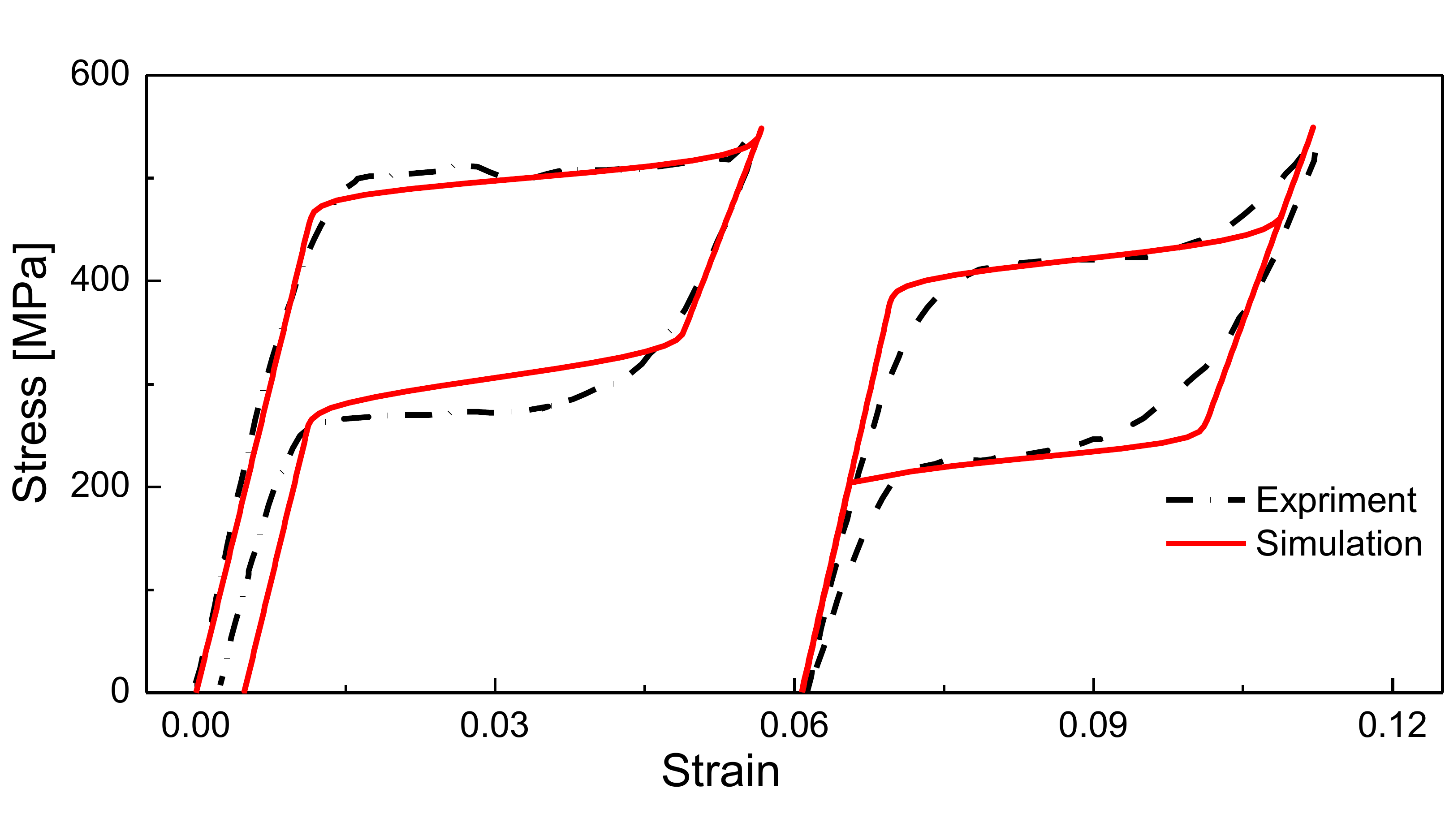}
	\end{center}
	\vspace*{-7mm}
	\caption{The $1^{st}$ and $50^{th}$ response of a NiTiCu SMA wire subjected to 50 unaxial tensile stress cycling. The experimental data used for comparison are referenced from \cite{strnadel2011}.}
	\label{fig:NiTi_1}
\end{figure}

\begin{figure}[h]
	\begin{center}
		\advance\leftskip-0.5cm
		\includegraphics[width=0.6\columnwidth]{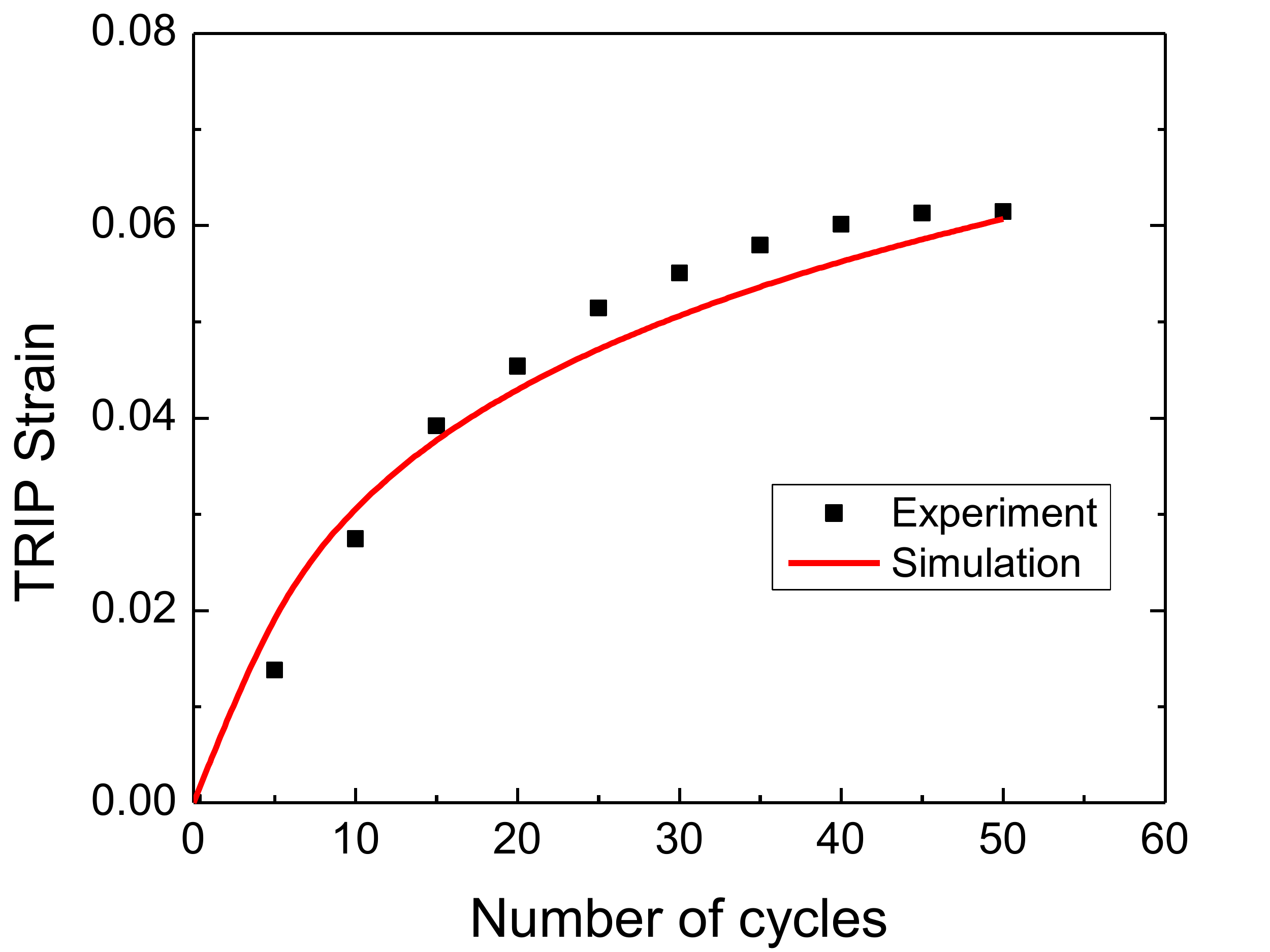}
	\end{center}
	\vspace*{-7mm}
	\caption{Accumulation of TRIP strain for the NiTiCu SMA wire subjected to 50 unaxial tensile stress cycling. The experimental data used for comparison are referenced from \cite{strnadel2011}.}
	\label {fig:NiTi_2}
\end{figure}

The cyclic stress-strain curve acquired by the proposed model is compared against the available experimental data reported in \cite{strnadel2011}. As shown in Fig. \ref{fig:NiTi_1}, the NiTiCu SMA accumulates a large amount of irrecoverable TRIP strain from the $1^{st}$ cycle to the $50^{th}$ cycle. In addition, it can be seen from Fig. \ref{fig:NiTi_2} that the TRIP strain grows drastically within the initial 30 loading cycles then stabilizes in a stationary increasing trend thereafter. The accumulated TRIP strain in the $1^{st}$ cycle is about 0.6\% and grows to around 6\% after 30 cycles. The predicted TRIP accumulation result is shown in good agreement with the experimental data. Apart from capturing the irrecoverable TRIP strain, the model is also able to capture the experimentally observed decreasing stress levels required to initiate the forward phase transformation. This modeling capability is achieved by the introduction of internal stress term into the model formulation. As the internal stress accumulates in an exponential manner described as evolution equation (\ref{eq:internal_stress}) when the SMA material is subjected to mechanically cycling, it decreases the stress required from outside to kick off the phase transformation. In summary, it is shown in this BVP that the proposed model is able to predict the TRIP strain accumulation during cyclic pseudoelastic loading, and also is capable to capture the decreasing stress level needed to initiate the phase transformations after the mechanically cycling.

\subsection{Uniaxial cyclic actuation loading case}\label{uniaxial_actuation}
In this section, two BVPs for SMAs with different material compositions, i.e., Ni$_{49.9}$Ti$_{50.1}$ (at.\%) and Ni$_{50.3}$Ti$_{29.7}$Hf$_{20}$ (at.\%), subjected to uniaxial actuation cycling are analyzed. The NiTi material is a classical type of SMA having critical phase transforming temperatures close to room temperature, while the NiTiHf belongs to the so-called high-temperature SMA category that can function under very extreme environments with their phase transformation temperatures around 100 $^{\circ}$C. These two BVPs are chosen to check the model's fidelity over multiple SMA material systems, also to check the proposed capability of capturing the load-free TWSME for thermomechanically trained SMAs. The load-free thermal cycling response of SMAs are examined and compared to available experimental data to validate such capability.

In the NiTi case, the SMA dogbone specimen was subjected to a 100 thermal cycling between 310 and 440 K under a constant stress 200 MPa. After the 100 training cycles, the bias load was removed, and the final thermal cycling was performed to check the TWSME response under load-free condition. The material parameters used here to simulate this experiment are listed in table \ref{tab:MP_NiTi}. Similar to the experimental procedure in the case of NiTi, the NiTiHf dogbone specimen was also loaded under 200 MPa and experienced a 100 thermal cycling between 310 and 580 K. The load-free TWSME was also checked after the thermomechanical training cycles. The material parameters used to simulate the NiTiHf actuation response are listed in table \ref{tab:MP_NiTiHf}. The experimental data used for comparison were initially reported in \cite{atli2015}.

\begin{figure}[H]
	\begin{center}
		\vspace*{-1cm}
		\advance\leftskip-1cm
		\includegraphics[width=0.9\columnwidth]{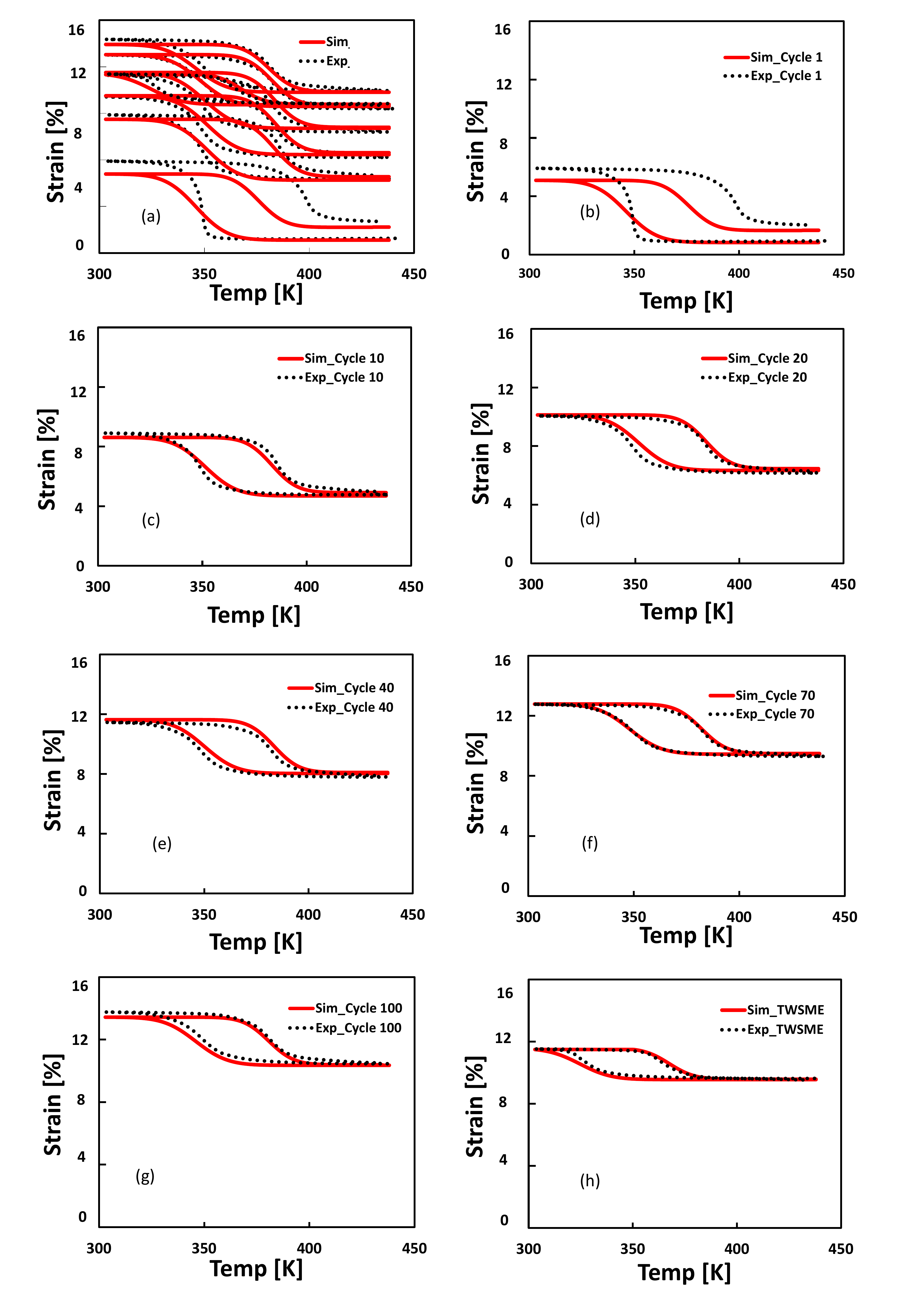}
	\end{center}
	\vspace*{-0.8cm}
	\caption{Selected thermal cycling and load-free TWSME response for the NiTi SMA under a 200 MPa constant load. (a) Combined, (b) Cycle 1, (c) Cycle 10, (f) Cycle 20, (e) Cycle 40, (f) Cycle 70, (g) Cycle 100, (h) TWSME cycle. The experimental data used for comparison are referenced from \cite{atli2015}. }
	\label{fig:NiTi_3}
\end{figure}

\begin{figure}[H]
	\begin{center}
		\vspace*{-1cm}
		\advance\leftskip-1cm
		\includegraphics[width=0.94\columnwidth]{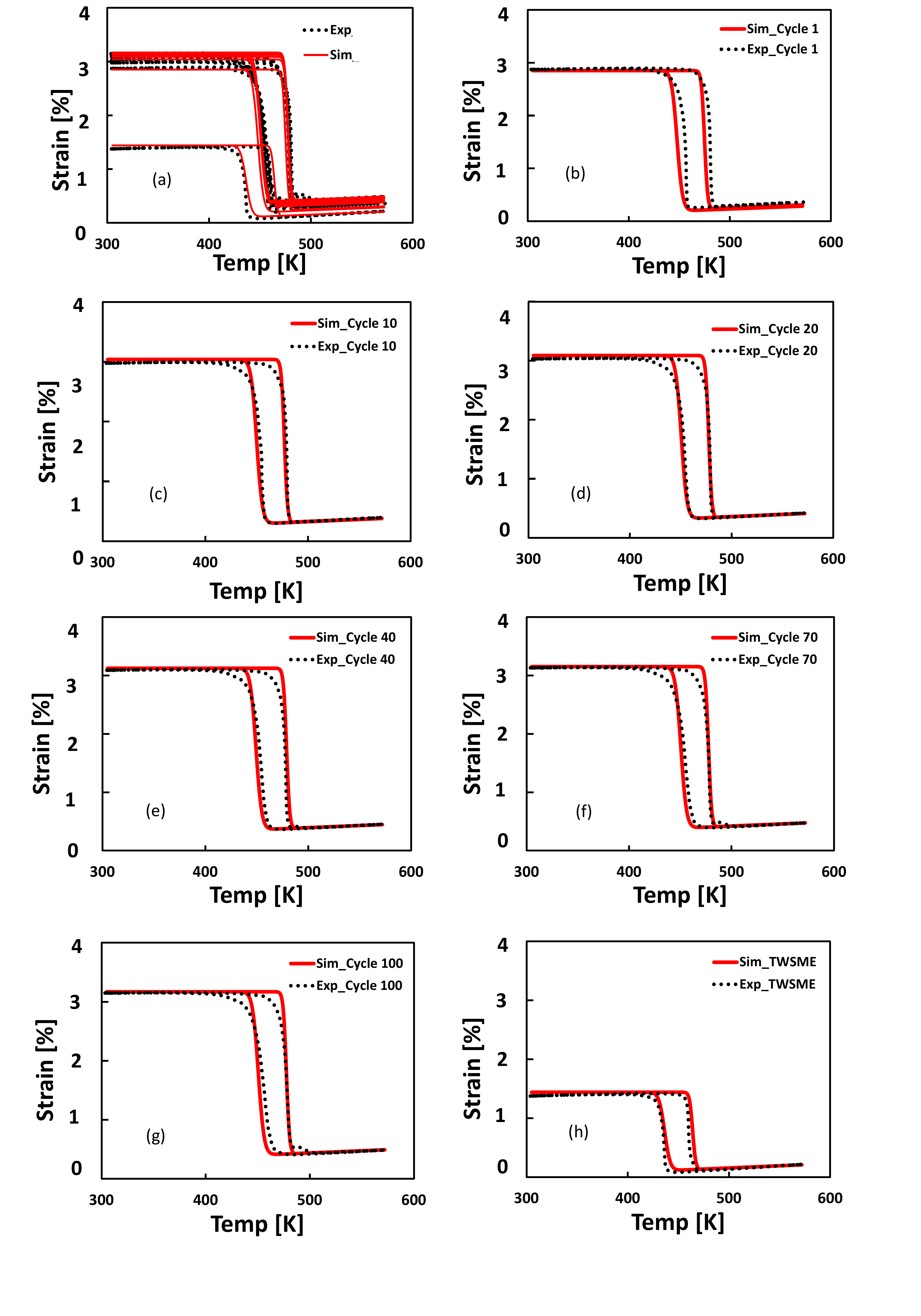}
	\end{center}
	\vspace*{-1.5cm}
	\caption{Selected thermal cycling and load-free TWSME responses for the NiTiHf SMA under a 200 MPa constant load. (a) Combined, (b) Cycle 1, (c) Cycle 10, (f) Cycle 20, (e) Cycle 40, (f) Cycle 70, (g) Cycle 100, (h) TWSME cycle. The experimental data used for comparison are referenced from \cite{atli2015}.}
	\label {fig:NiTiHf_1}
\end{figure}

The simulated cyclic actuation response by the proposed model are summarized in Fig. \ref{fig:NiTi_3} for NiTi SMA and in Fig. \ref{fig:NiTiHf_1} for NiTiHf SMA. In the NiTi case, the results for selected training cycle 1, cycle 10, cycle 20, cycle 40, cycle 70, and cycle 100 are plotted and compared to the experiment results. It is shown that the simulated transformation characteristics in these selected training cycles, including transformation temperatures, transformation train magnitude, and TRIP strain accumulation, are in good agreement with the experiment data. More specifically, the NiTi SMA accumulated about 11\% TRIP strain after the 100 thermal training cycle. Additionally, the simulated load-free TWSME curve for NiTi SMA also correlates well with the experimental one. In the case of NiTIHf, it can be seen that the high-temperature SMA material system shows a quite different actuation response compared to that of NiTi. Specifically, the phase transformation temperatures are much higher, and much less TRIP strain is accumulated in the end. Although those characteristics are quite different compared to NiTi, the simulated results also agree well with the experimental data. The comparison between simulation results and experimental data demonstrates that the proposed model is enabled to capture the load-free TWSME response for thermomechanically trained SMAs, and also the fidelity of the proposed modeling capabilities over multiple SMA material systems.

\begin{table}[h] 
	\centering
	\caption{Model parameters used for NiTi SMA under uniaxial cyclic actuation loading.}
	\renewcommand{\arraystretch}{0.7}
	\begin{tabular}{c|lr|ll} \toprule
		Type                         &Parameter                        & Value                                   &Parameter            & Value  \\                                       \midrule
		&$E_A$                       & 24.15   [GPa]                   & $C_A$                                   & 15  [MPa/K]\\
		&$E_M$                       & 24.15   [GPa]                   & $C_M$                                   & 8  [MPa/K]\\
		Key material parameters       &$\nu_A=\nu_M$                   & 0.33~~~~~~~~                              & $M_s$                & 330  [K]\\
		12                            &$\alpha_A=\alpha_M$             & 1.0$\times$10$^{-5}$ [K$^{-1}$]                           & $M_f$                &300  [K]\\
		& $ H^\text{max}_i$          & 4\%                            & $A_s$                 &351  [K]\\
		& $ H^\text{max}_f$          & 3.17\%                            & $A_f $                &375  [K] \\
		& $ H^\text{min} $           & 0\%                         & $k_t$                 &  0.045 [1/MPa]\\    
		& $\uptau_{\text{crit}}$      &&&                             \\   \midrule
		Smooth hardening parameters 		&  $n_1$         &   0.5                &  $n_3$             & 0.5 \\
		4							   &  $n_2$         &   0.5                &  $n_4$        	 & 0.5 \\  \midrule
		
		&${\sigma}_b$                        & 80    [MPa]                      & $\lambda_1$                & 0.1 \\
		TRIP parameters               &$C_1^p$                      & 2.6$\times$10$^{-2}$                     &                  &   \\
		4                     &$C_2^p$                      & 0.18                                 &                 &  \\                              
		\bottomrule
	\end{tabular}
	\label{tab:MP_NiTi}
\end{table}

\begin{table}[H] 
	\centering
	\caption{Model parameters used for NiTiHf SMA under  uniaxial cyclic actuation loading.}
	\renewcommand{\arraystretch}{0.7}
	\begin{tabular}{c|lr|ll} \toprule
		Type                         &Parameter                        & Value                                   &Parameter            & Value  \\                                       \midrule
		&$E_A$                       & 70   [GPa]                   & $C_A$                                   & 12.5  [MPa/K]\\
		&$E_M$                       & 70   [GPa]                   & $C_M$                                   & 11.9  [MPa/K]\\
		Key material parameters       &$\nu_A=\nu_M$                   & 0.33~~~~~~~~                              & $M_s$                & 441  [K]\\
		12                            &$\alpha_A=\alpha_M$             & 1.0$\times$10$^{-5}$ [K$^{-1}$]           & $M_f$                & 430  [K]\\
		& $ H^\text{max}_i$              & 3.15\%                            & $A_s$                 & 460  [K]\\
		& $ H^\text{max}_f$              & 3.15\%                            & $A_f $                & 466  [K] \\
		& $ H^\text{min}$                & 0.7\%                         & $k_t$                 &  0.007 [1/MPa] \\   
		& $\uptau_{\text{crit}}$       & 0 [MPa]                              &                       &                             \\   \midrule
		
		Smooth hardening parameters 		&  $n_1$         &   0.06                &  $n_3$             & 0.06 \\
		4							        &  $n_2$         &   0.06                &  $n_4$        	 & 0.06 \\  \midrule
		
		&${\sigma}_b$                        & 35    [MPa]                      & $\lambda_1$                & 0.1 \\
		TRIP parameters               &$C_1^p$                      & 5$\times$10$^{-4}$                     &                  &   \\
		4                     &$C_2^p$                      & 0.3                                 &                 &  \\                              
		\bottomrule
	\end{tabular}
	\label{tab:MP_NiTiHf}
\end{table}

\newpage

\begin{figure}[H]
	\begin{center}
		\advance\leftskip-1cm
		\includegraphics[width=1.1\columnwidth]{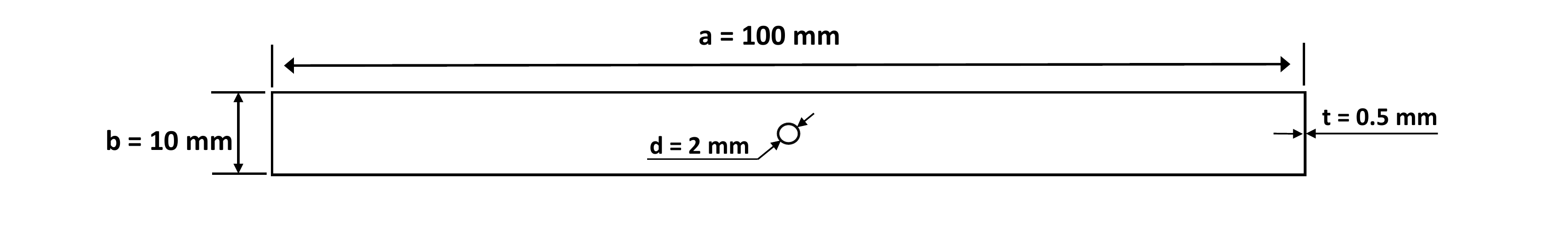}
	\end{center}
	\vspace*{-0.8cm}
	\caption{The geometry of the Ni$_{55}$Ti$_{45}$ (at.\%) plate actuator with a centric hole. }
	\label{fig:PlateGeom}
\end{figure}

\begin{table}[H] 
	\centering
	\caption{Model parameters used for the Ni$_{55}$Ti$_{45}$ (at.\%) plate actuator.}
	\renewcommand{\arraystretch}{0.7}
	\begin{tabular}{c|lr|ll} \toprule
		Type                         &Parameter                        & Value                                   &Parameter            & Value  \\                                       \midrule
		&$E_A$                            & 70   [GPa]                              & $C_A$                & 22  [MPa/K]\\
		&$E_M$                            & 70   [GPa]                              & $C_M$               & 22  [MPa/K]\\
		Key material parameters       &$\nu_A=\nu_M$                          & 0.33~~~~~~~~                              & $M_s$                & 318  [K]\\
		12                      &$\alpha_A=\alpha_M$    &    1.0$\times$10$^{-5}$ [K$^{-1}$]                           & $M_f$                &298  [K]\\
		& $ H^\text{max}_i$              & 1.5\%                            & $A_s$                 &  332  [K]\\
		& $ H^\text{max}_f$              & 1.5\%                              & $A_f $                &  352  [K] \\
		& $ H^\text{min}$            & 0\%                            & $k_t$                 &  0.01 [1/MPa] \\                                  
		& $\uptau_{\text{crit}}$      &0 [MPa]&&                             \\    \midrule
		
		Smooth hardening parameters 		&  $n_1$         &   0.5                &  $n_3$             & 0.5 \\
		4							   &  $n_2$         &   0.5                &  $n_4$        	 & 0.5  \\   \midrule
		
		&${\sigma}_b$                        & 0    [MPa]                      & $\lambda_1$                & N/A \\
		TRIP parameters               &$C_1^p$                      & 5.73$\times$10$^{-3}$                     &                  &   \\
		4                     &$C_2^p$                      & 1.972                                 &                 &  \\                              
		\bottomrule
	\end{tabular}
	\label{tab:MP_Plate}
\end{table}

\subsection{Plate actuator with a centric hole under cyclic actuation loading}
After the analysis of SMAs under uniaxial loading condition, this part analyzes the BVP when SMAs are subjected to multiaxial stress conditions. As it was mentioned in the introduction, it is imperative to understand the cyclic response of SMAs under the multiaxial loading conditions, as the majority of applications require the functionality of SMA-based components under multiaxial stress state that are caused by geometry discontinuities, such as notches and holes coming from installment requirements. It has been reported that TRIP strain evolves differently in the multiaxial stress state compared to that in the uniaxial one. In order to study the effect of the stress multiaxiality, a plate actuator with a centric hole made from Ni$_{55}$Ti$_{45}$ (at.\%) subjected to cyclic actuation loading is chosen as the BVP to be investigated. The geometry of the plate actuator is shown in Fig. \ref{fig:PlateGeom}, which has 100 mm in the length, 10 mm in the width, and 0.5 mm in the thickness. The centric hole has a diameter of 2 mm. In the experiment, a nominal load of 136 MPa was applied in the longitudinal direction of the plate, thereafter it was subjected to a thermal cycling between 280 and 400 K for 100 cycles while the bias load was maintained constant. The TRIP strain in the longitudinal direction was recorded at the end of each thermal cycling by using the Digital Correction Image (DIC) technique. A simulation was performed by using the presented model wherein the loading conditions are the modeled the same as the experiment. The material parameters used in this simulation are listed in table \ref{tab:MP_Plate}. 

\begin{figure}[H]
	\begin{center}
		\advance\leftskip-0.6cm
		\includegraphics[width=0.9\columnwidth]{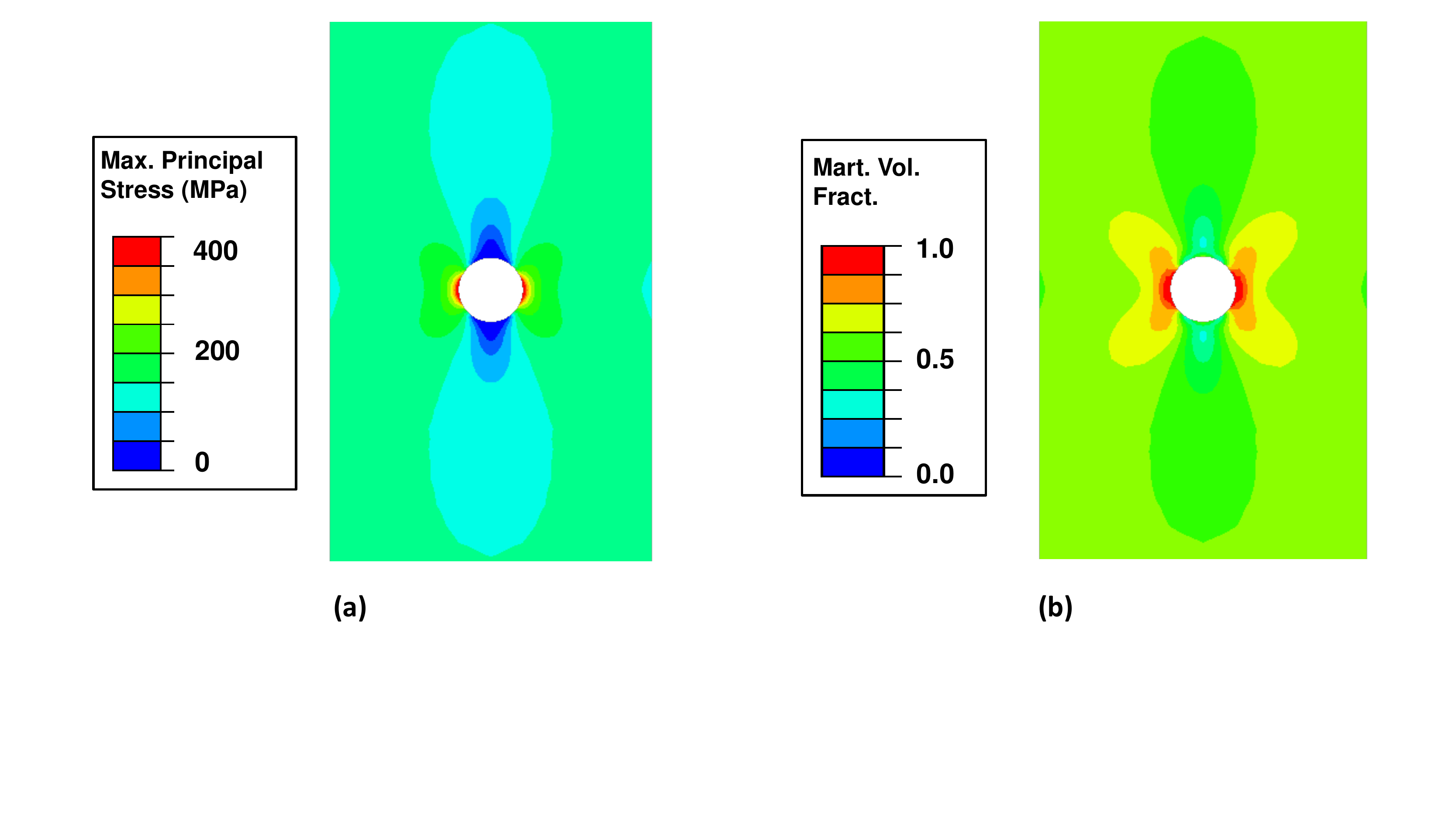}
	\end{center}
	\vspace*{-1.5cm}
	\caption{The simulated stress distribution and martensitic volume fraction contour for the Ni$_{55}$Ti$_{45}$ (at.\%) plate actuator. (a) Maximum principal stress, (b) martensitic volume fraction.}
	\label{fig:PlateContour}
\end{figure}
\begin{figure}[H]
	\begin{center}
		\advance\leftskip-0.7cm
		\includegraphics[width=1.05\columnwidth]{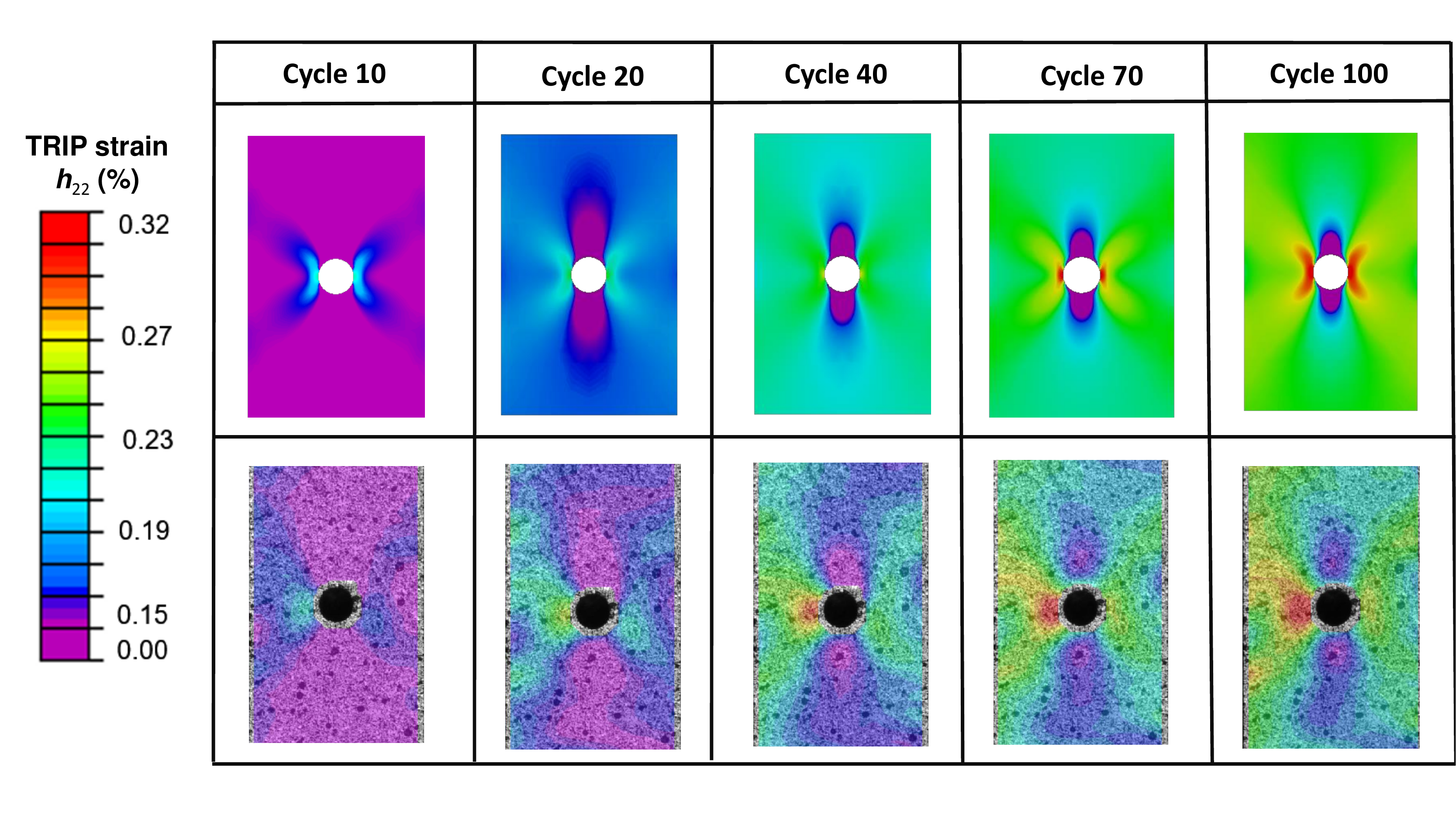}
	\end{center}
	\vspace*{-0.5cm}
	\caption{TRIP strain evolution contour for the Ni$_{55}$Ti$_{45}$ (at.\%) plate actuator subjected to cyclic thermomehcanical loading. The first row in this figure is simulation while the second row is experimental DIC result. The experimental data is referenced from \cite{wheeler2017actuation}.}
	\label{fig:Plate_compar}
\end{figure}

As it can be seen from the principal stress contour in Fig.~\ref{fig:PlateContour}(a), the non-uniform multiaxial stress field is caused due to the presence of the hole, a much larger stress field in the designated red region while a much smaller stress field in the rest part. As a result of such stress multiaxiality, the phase transformation of the plate actuator is largely redistributed. As it is shown Fig. \ref{fig:PlateContour} (b), the stress concentrated part starts the phase transformation earlier while in the less stressed part it is later initiated. Furthermore, refer to Fig. \ref{fig:Plate_compar} for the TRIP strain accumulation results, the experimental DIC result is shown in the bottom row while the prediction is shown in the upper row of the figure. The experimental results show that the multiaxial stress state has a significant effect on the TRIP strain evolution from cycle to cycle. Specifically, the TRIP strain accumulates much faster in the stress concentrated part compared to the less stress concentrated part. In cycle 10, There is around 0.17\% TRIP strain accumulated in the stress concentration part while the rest of the plate actuator is much less than 0.1\%. In cycle 100, there is 0.32\% TRIP strain generated in the red area while the less stressed part accumulates around 0.15\%, and the rest part of the plate under uniform stress field accumulates around 0.25\%.  The observation on the experimental results indicated that the generation of TRIP strain is highly stress-dependent. Although this feature seems very intrinsic to model, the proposed model, however, did a quite good job on predicting that. The simulation results not only well captured the overall distribution shape for the accumulated TRIP strain, the magnitude in the stress concentrated and less stressed regions are also in good agreement with the experiments. In this BVP, the proposed model is demonstrated to have the capability to predict the highly stress-dependent TRIP strain generation when SMAs are subjected to a multiaxial stress state. 



%

\section{CONCLUSIONS}\label{conc}
In this work, a three-dimensional finite strain constitutive model for SMAs considering the stress-dependent TRIP evolution under multiaixal stress state, as well as the TWSME at load-free condition is proposed. This model is developed based upon the early work by \cite{lagoudas2012}, and largely inspired by its consideration of TRIP \citep{Bo1999_TRIP_3,lagoudas2004TRIP}, as well as its recent extension into large deformation framework \citep{xu_2019sms}. By consideration of the martensitic volume fraction, transformation strain, internal stress, and TRIP strain tensors as internal state variables, the model is enabled to capture the following important characteristics for polycrystalline SMAs. (1) The model formulated based on a finite deformation theory is inherently able to capture the large strains and rotations exhibited by SMAs under cyclic thermomechanical loading. (2) Through the introduction of an accumulating internal stress tensor, the model is capable of predicting the TWSME for thermomechanically trained SMAs under load-free conditions, and also capturing the decreasing stress level to initiate the phase transformation during pseudoelastic cycling. (3) By proposing a new TRIP strain evolution law, the model is able to predict the stress-dependent  TRIP strain generation when SMAs are subjected to a multiaxial stress state. A detailed implementation procedure of the proposed model is presented through a user-defined material subroutine within a finite element framework allowing for solving complex BVPs, and a comprehensive instruction on calibrating the model parameters as well as the derivation of continuum tangent stiffness matrix are also provided. The proposed modeling capabilities have been validated through both unaxial and multiaxial BVPs for a wide range of SMA material systems under different thermomehcnaical loading paths, which demonstrates the presented model's fidelity as an efficient design and analysis tool for the future applications of SMA-based multifunctional components.

\section*{ACKNOWLEDGMENT}
The authors would like to acknowledge the financial support provided by the Qatar National Research Fund (QNRF) under the grant number: NPRP 7-032-2-016, and the National Aeronautics and Space Administration (NASA) through the University Leadership Initiative (ULI) program under the grant number: NNX17AJ96A. The authors would also like to thank Drs. Ibrahim Karaman, Robert Wheeler, Kadri Can Atli, and Mr. Glen Bigelow for providing the invaluable experimental data for model validation. At last, the authors would like to thank Dr. Anargyros Karakalas to proof reading the manuscript and provide his invaluable comments.

\section*{DATA AVAILABILITY}
The user-defined material subroutine (in the form of Abaqus UMAT) described in this paper for the proposed SMA constitutive model considering TRIP and TWSME  is available to access from GitHub: https://github.com/Aero-tomato/SMA-UMAT.

\bibliography{myarticle}
\newpage
\appendix
\section{Continuum tangent stiffness and thermal moduli}\label{sec:Jacobian}
In order for the displacement-based Finite element solver to attain a converged solution for the global equilibrium equations, the so-called tangent stiffness matrix must be defined in the UMAT. In general, the aforementioned matrix can be derived using either the continuum or the time-discretized equations of stress and transformation function. The tangent stiffness matrix derived using the time-discretized equations is commonly referred as consistent stiffness matrix, which is the consistent one used in the UMAT framework, and therefore leads to a quadratic convergence rate in the Newton-Raphson process within the finite element solver. The tangent stiffness matrix, derived using the continuum equations, is called the continuum stiffness matrix and it may lead to a slower convergence rate in comparison to the time discretized form, especially when large analysis time steps are used. However, it should be noted that both approaches will lead to the same solution once the Newton-Raphson process converges.  An extensive discussion about this matter can be found in \cite{lagoudas2008}. In the current work, the continuum tangent matrix is the opted one due to the simplicity of its derivations. Although we acknowledge the aforementioned potential disadvantages, that could be addressed in a future work to maximize the algorithm performance.

A detailed derivation of the continuum tangent stiffness matrix is provided in the following part. For the sake of completeness, the presented derivation has also included the derivation of the tangent thermal moduli. Although it should be noted that the tangent thermal moduli is required to be defined in the UMAT, only if the coupling between the thermal and mechanical equilibrium equations is considered, which has not been considered in the current work. In general, these continuum tangent tensors can be formulated as shown in equation (\ref{eq:Jacobian}), where $\mathcal{L}$ is called the tangent stiffness matrix and $\Theta$ is the tangent thermal moduli. Be noted that the Vogit notation of tensors is used in the actual implementation.
\begin{equation}\label{eq:Jacobian}
\mathring{\bm{\uptau}}  =  {\mathcal{L}}\mathring{\mathbf h} 
+ \Theta \dot T
\end{equation}
Applying the logarithmic rate on constitutive equation (\ref{eq:h_Cons_f}) yields, 
\begin{equation}\label{eq:rate_cons}
\mathring{\bm{\uptau}}  = {\mathcal{C}}~[\mathring {\mathbf h}  -  \bm\alpha\dot T - ({\Delta\mathcal{S}}\bm{\uptau} +  \Delta{\bm\alpha}\Delta T + \mathbf\Lambda +\mathbf\Lambda^{tp})\dot \xi~]
\end{equation}
Taking the chain rule differentiation on the transformation function equation (\ref{eq:Transfor_Fun}) gives,
\begin{equation}\label{eq:rate_TransFun}
\dot{\Phi}  = \partial_{\bm{\uptau}}\Phi:\mathring{\bm\uptau} + \partial_{T}\Phi\dot{T} + \partial_{\xi}\Phi\dot{\xi} = 0  
\end{equation}
Substitute equation (\ref{eq:rate_cons}) into equation (\ref{eq:rate_TransFun}) to cancel $\mathring{\bm{\uptau}}$ and solve it for $\dot \xi$, the following expression for $\dot \xi$ can be obtained,
	\begin{equation}\label{eq:rate_xi}
	\dot{\xi}  = -\dfrac{\partial_{\bm\uptau}\Phi: \mathcal{C} \mathring {\mathbf h} +(\partial_{T}\Phi- \partial_{\bm\uptau}\Phi: \mathcal{C} \bm \alpha)\dot T}{\partial_{\xi}\Phi- \partial_{\bm\uptau}\Phi: \mathcal{C} (\Delta\mathcal{S}\bm\uptau+\bm\Lambda+\bm\Lambda^{tp}+\Delta{\bm\alpha}\Delta T)} 
	\end{equation}

As the phase difference of thermal expansion coefficients $\Delta \bm{\alpha}$ is quite small usually, it is neglected in the derivation for the sake of brevity. The final explicit expression in the form as equation (\ref{eq:Jacobian}) can be obtained as follows by substitution of equation (\ref{eq:rate_xi}) into the rate form constitutive equation (\ref{eq:rate_cons}), 
\begin{equation}\label{eq:Jacobian_exp}
\begin{aligned}
\mathring{\bm{\uptau}}  \quad = \quad &\Big[ \mathcal{C}+\dfrac{[\mathcal{C}({\Delta\mathcal{S}}\bm{\uptau}+ \bm\Lambda +\mathbf\Lambda^{tp})] \otimes  [\mathcal{C} \partial_{\bm\uptau}\Phi]}{\partial_{\xi}\Phi- \partial_{\bm\uptau}\Phi: \mathcal{C}(\Delta\mathcal{S}\bm\uptau+\bm\Lambda+\mathbf\Lambda^{tp})} \Big] \mathring{\mathbf{h}} 
\quad + \\ & \Big[- \mathcal{C}\bm\alpha + \dfrac{ \mathcal{C}({\Delta\mathcal{S}}\bm{\uptau}+ \bm\Lambda +\mathbf\Lambda^{tp})  (\partial_{T}\Phi - \partial_{\bm\uptau}\Phi: \mathcal{C}\bm\alpha )}
{\partial_{\xi}\Phi- \partial_{\bm\uptau}\Phi: \mathcal{C}(\Delta\mathcal{S}\bm\uptau+\bm\Lambda +\mathbf\Lambda^{tp})}  \Big] \dot T 
\end{aligned}
\end{equation}
in which the continuum tangent stiffness matrix ${\mathcal{L}}$ is, 
\begin{equation}
{\mathcal{L}}=\mathcal{C}+\dfrac{[\mathcal{C}({\Delta\mathcal{S}}\bm{\uptau}+ \bm\Lambda +\mathbf\Lambda^{tp})] \otimes  [\mathcal{C} \partial_{\bm\uptau}\Phi]}{\partial_{\xi}\Phi- \partial_{\bm\uptau}\Phi: \mathcal{C}(\Delta\mathcal{S}\bm\uptau+\bm\Lambda+\mathbf\Lambda^{tp})}
\end{equation}
and the continuum thermal moduli $\Theta$ is, 
\begin{equation}
\Theta = 
\dfrac{ \mathcal{C}({\Delta\mathcal{S}}\bm{\uptau}+ \bm\Lambda +\mathbf\Lambda^{tp})  (\partial_{T}\Phi - \partial_{\bm\uptau}\Phi: \mathcal{C}\bm\alpha )}
{\partial_{\xi}\Phi- \partial_{\bm\uptau}\Phi: \mathcal{C}(\Delta\mathcal{S}\bm\uptau+\bm\Lambda +\mathbf\Lambda^{tp})} - \mathcal{C}\bm\alpha
\end{equation}

In order to fully determine the explicit expressions for $\mathcal{L}$ and ${\Theta}$ during the implementation, the following terms $\partial_{\bm\uptau}\Phi$, $\partial_{\xi}\Phi$, $\partial_{T}\Phi$ used in above equations can be calculated by using the symbolic calculation toolbox provided in MATLAB.

\section{Model calibration}\label{sec:calibration}
This section presents a detailed instruction on how to identify all the model parameters from a set of calibration tests. In general, those parameters can be categorized into three major groups, {i.e.}, the key material parameters, smooth hardening parameters, TRIP and internal stresses parameters. Note that the strain measure used here is  logarithmic (or true) strain rather than engineering (or infinitesimal) strain.

\begin{figure}[H]
	\advance\leftskip-0.5cm
	\subfigure[]{\includegraphics[width=0.55\columnwidth]{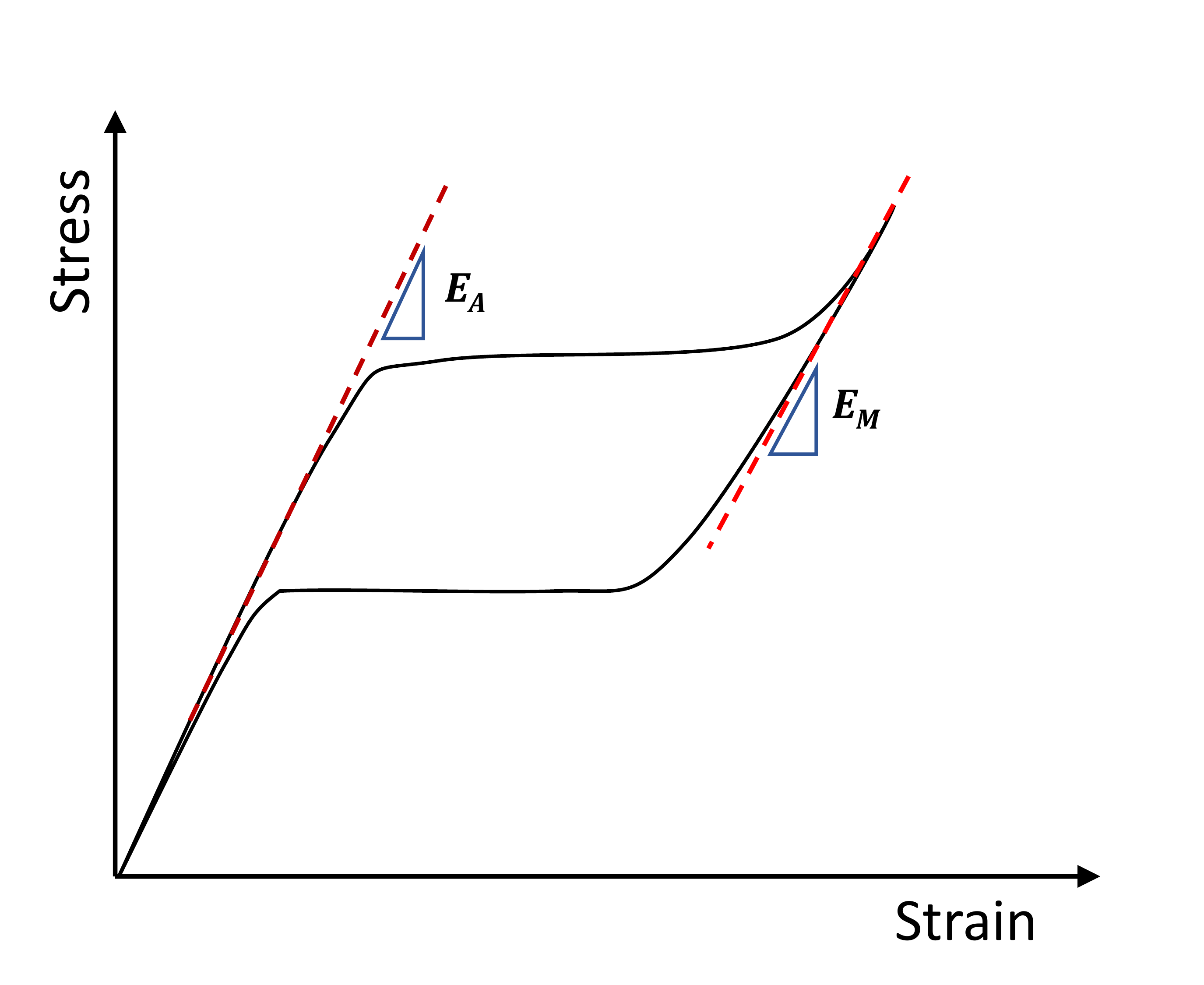}
		\label{fig:Uniaxial}}
	\subfigure[]{\includegraphics[width=0.55\columnwidth]{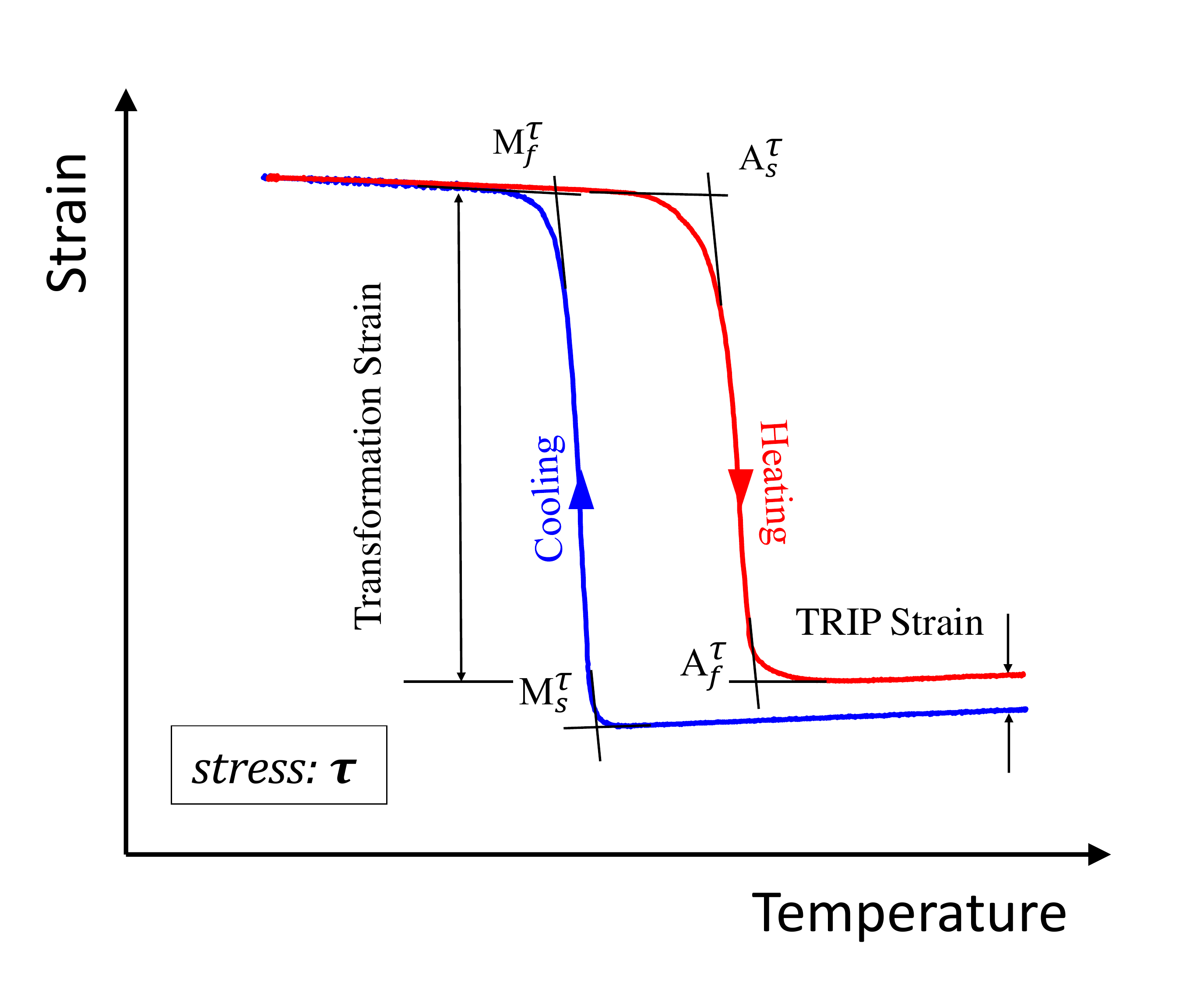}
		\label{fig:Actuation}}
	\vspace*{-2mm}
	\caption{Experiments utilized for the calibration of model parameters. (a) Uniaxial pseudoelastic experiment used for the calibration of elastic modulus of austenite and martensit. (b) Uniaxial actuation experiments under three different stress levels used for the calibration of phase diagram related model parameters.}
	\label{fig:cali_1}
\end{figure}

\begin{figure}[H]
	\advance\leftskip-0.5cm
	\subfigure[]{\includegraphics[width=0.55\columnwidth]{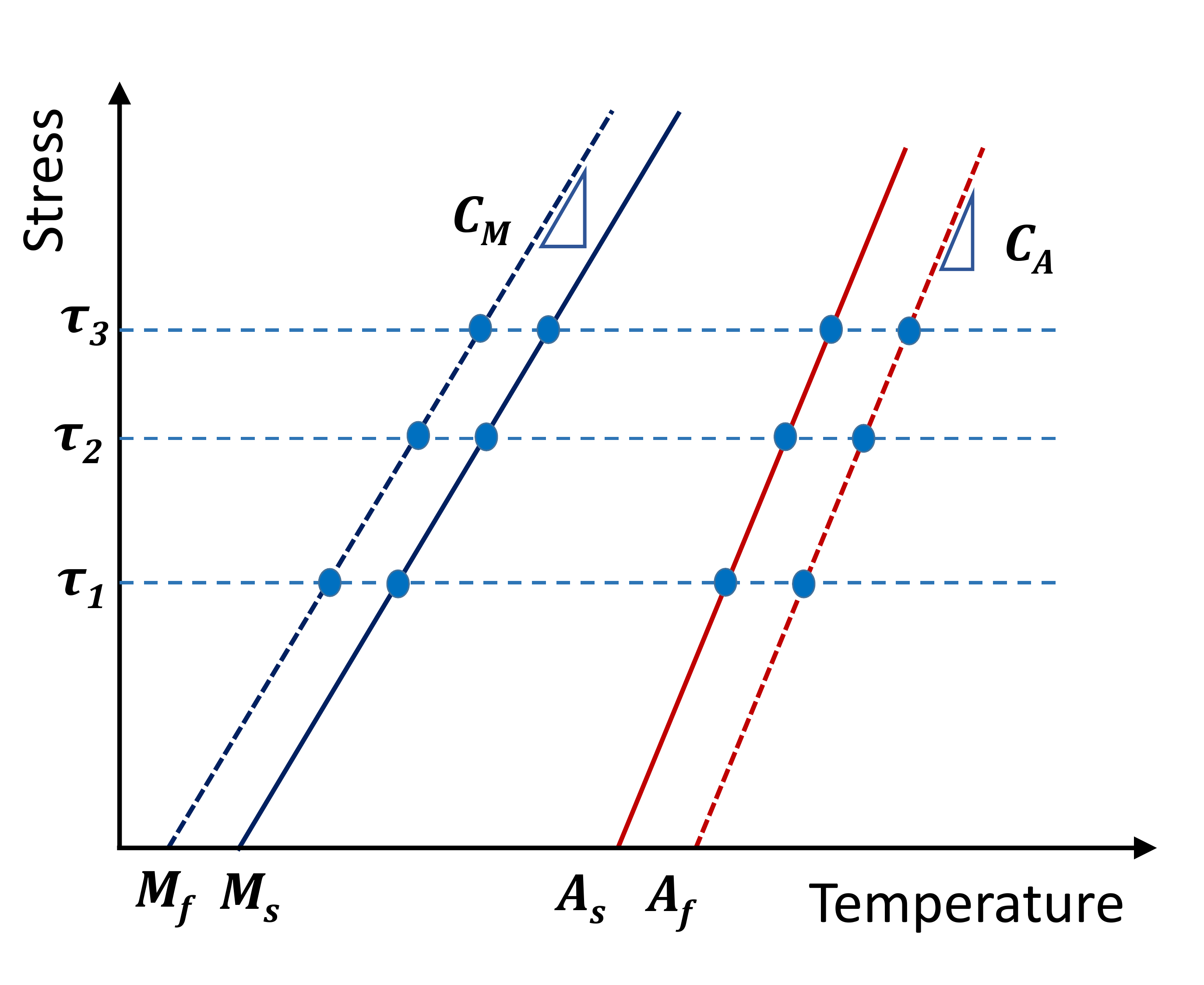}
		\label{fig:PhaseDiagram}}
	\subfigure[]{\includegraphics[width=0.55\columnwidth]{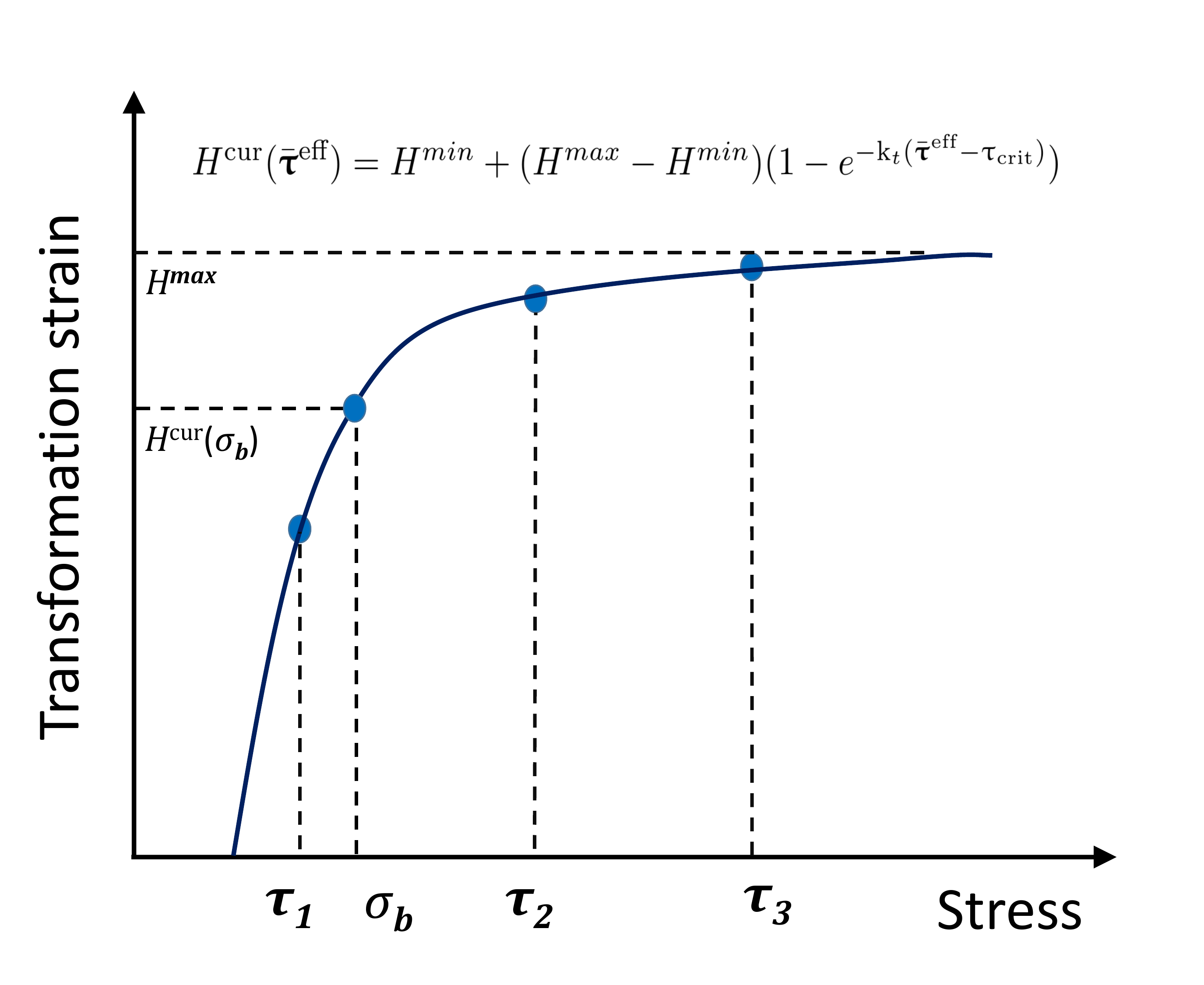}
		\label{fig:Hcur}}
	\vspace*{-2mm}
	\caption{Calibration of model parameters. (a) Phase diagram. (b) $H^{\textit{cur}}$ curve.}
	\label{fig:cali_2}
\end{figure}

First, material constants such as the elastic modulus $(E_A, E_M)$ of austenite and martensite can be obtained by calculating the slopes at martensitic phase and austenite phase from uniaxial mechanical loading-unloading experiment, i.e., the pseudoelastic response as shown in Fig. \ref{fig:Uniaxial}. Poisson's ratios $\nu_A$ and $\nu_B$ are usually assumed as $0.33$ for metallic materials. In order to construct the phase diagram, thermal cycling of the material under constant uniaxial tensile stress, i.e., actuation response as shown in Fig. \ref{fig:Actuation}, is performed, by which critical transformation temperatures ($M_s^{\uptau}, M_f^{\uptau}, A_s^{\uptau}, A_f^{\uptau}$) and transformation strain $H^{cur}(\uptau)$ for a specific applied stress level ${\uptau}$ are measured. Such experiments are performed at three different stress levels ($\uptau_1, \uptau_2, \uptau_3$). By using these collected temperature and transformation strain information, the phase diagram and the $H^{cur}$ curve can be constructed, see Fig. \ref{fig:cali_2}. Therefore, phase diagram related stress influence coefficients ($C_A, C_M$), phase transformation temperatures $(M_s,M_f,A_s,A_f)$ at zero stresses, are determined. In addition, model parameters related to $H^{cur}$ curve $(H^{min},H^{max}, k_t, \uptau_{\text{crit}})$ are also obtained. Secondly, the smooth hardening related coefficients $(n_1, n_2, n_3, n_4)$ are determined by best matching the smoothness of response curve corner at the initiation and completion during phase transformation. The thermal expansion coefficients, usually assumed there is no phase differences $\alpha_A=\alpha_M$, can also be determined through an actuation response.

\begin{figure}[t]
	\advance\leftskip-0.5cm
	\subfigure[]{\includegraphics[width=0.55\columnwidth]{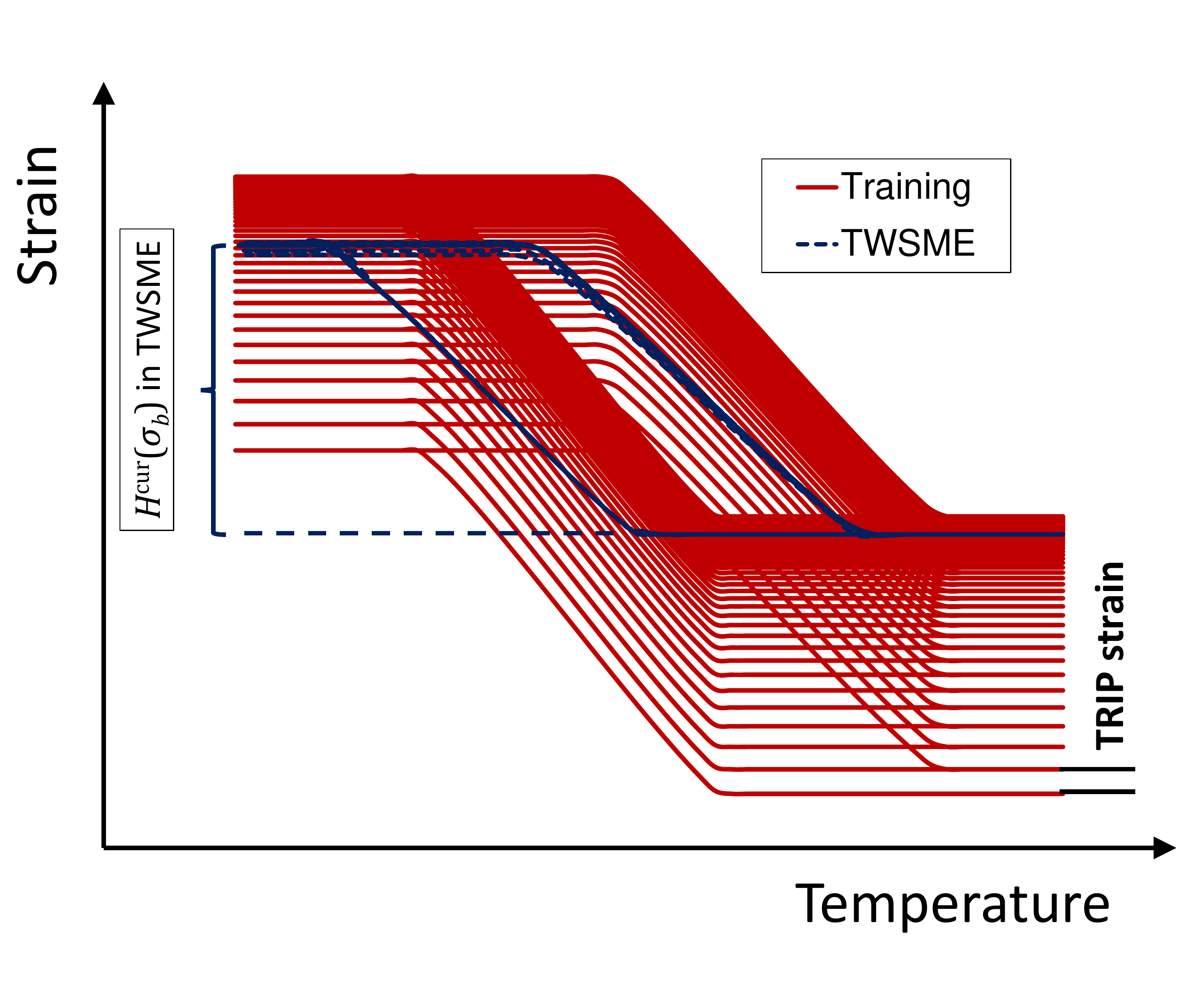}

		\label{fig:TRIP}}
	\subfigure[]{\includegraphics[width=0.55\columnwidth]{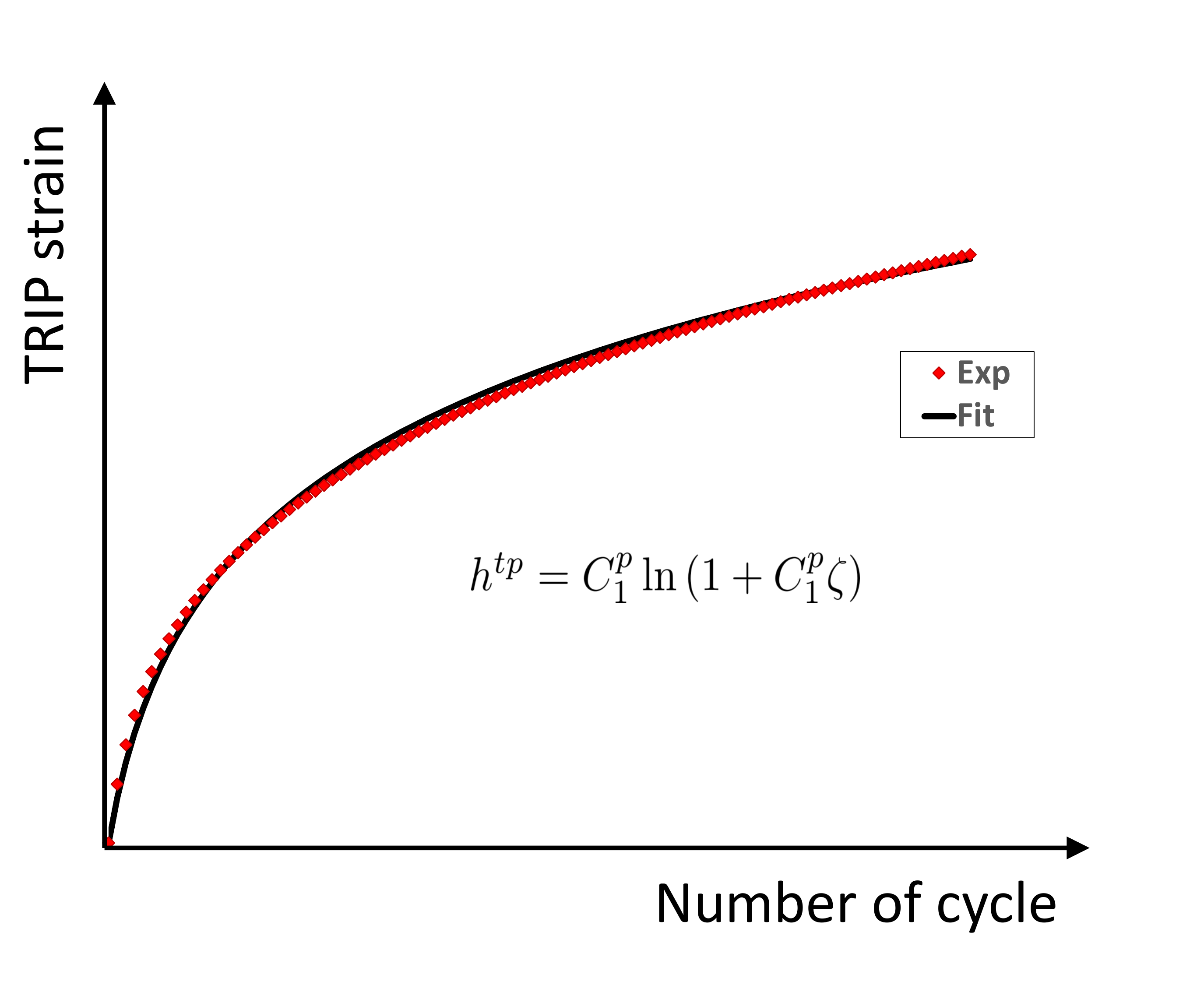}
		\label{fig:TRIP_Cali}}
	\vspace*{-2mm}
	\caption{Calibration of TRIP and internal stress related model parameters. (a) Cyclic actuation response with TWSME curve at load-free condition. (b) Calibration of TRIP related model parameters using TRIP strain versus number of loading cycle.}
	\label{fig:cali_3}
\end{figure}

In order to calibrate the TRIP related model parameters $(C_1^p, C_2^p)$, the thermal cycling of SMAs, cyclic actuation response shown as Fig. \ref{fig:TRIP}, subjected to a constant stress level is needed, in which the stress magnitude should exceed a saturation point beyond which no more transformation strain is generated even the stress further increases. From such experiment, the collected TRIP strain with respect to the number of loading cycle are used to calibrate the TRIP parameters. Integrating the rate form TRIP evolution equation (\ref{eq:Dir_plastic}) can obtain the algebraic form equation (\ref{eq:htp1}). Here the ratio term ${H^{cur}}/{H^{max}}$ is to incorporate stress-dependency effect on $C_1^p$ when SMAs are subject to multiaxial stress state. 
\begin{equation}\label{eq:htp1}
h^{tp}=\dfrac{H^{cur}}{H^{max}}C_1^p \ln(1+C_2^p \zeta^d) 
\end{equation} 

Under the condition that the material is fully transformed into oriented martensite phase at a saturated stress value, ${H^{cur}}/{H^{max}}=1$ and the accumulation of orientated martensitic volume fraction is reduced to the total one, i.e., $\zeta^d=\zeta$. Therefore, equation (\ref{eq:htp1}) can be further reduced into equation (\ref{eq:htp2}). TRIP related model parameters ($C_1^p, C_2^p$) can thus be calibrated by best fitting the TRIP strain evolution curve using the logarithmic function in equation (\ref{eq:htp2}) as shown in Fig. \ref{fig:TRIP_Cali}.
\begin{equation}\label{eq:htp2}
h^{tp}=C_1^p \ln(1+C_2^p \zeta) 
\end{equation}

\vspace{-0.5cm}

%

Next, internal stress related model parameters $(\sigma_b,\lambda_1)$ can be calibrated from the same thermal cycling experiment as shown in Fig.\ref{fig:TRIP} with additional information on TWSME response. As the internal stress is responsible for the generation of TWSME at load-free condition, it can be inversely determined by using the TWSME curve. Specifically, the magnitude of transformation strain $H^{cur}(\sigma_b)$ for the TWSME curve can be measured as shown in Fig.\ref{fig:TRIP}. Based on the $H^{cur}$ curve from Fig. \ref{fig:Hcur}, in order to produce this much transformation strain, the effective stress value of $\uptau^{\text{eff}}$ can be calculated. Also, it is straightforward to get $\uptau^{\text{eff}}=\sigma_b$ under load-free condition. Thus the value of $\sigma_b$ can be obtained by using the following derivation,
\begin{equation}
H^{\text{cur}}({\sigma_b})=H^{min}+(H^{max}-H^{min})(1-e^{-{k}_t (\sigma_b-\uptau_{\text{crit}})})	
\end{equation}
The explicit calculation of $\sigma_b$ is as follow,
\begin{equation}
\sigma_b=\uptau_{\text{crit}}+\frac{1}{{k}_t}\ln\big(\frac{H^{max}-H^{min}}{H^{max}-H^{cur}({\sigma_b})}\big)
\end{equation}

The additional model parameter $\lambda_1$ is a curve fitting parameter which controls how fast the quantities, such as the internal stress or $H^{max}$, evolve into their saturated values. It can be determined by best fitting the internal stress evolution law (\ref{eq:internal_stress}) or $H^{max}$ degradation law (\ref{eq:Hmax}) based on their evolution information. However, such information is often not easy to be measured, and $\lambda_1 = 0.01 \sim 1$ is an acceptable range.

There are also seven intermediate model parameters $(\rho_0\Delta s_0, \rho_0\Delta u_0, a_1, a_2, a_3, Y_0, D)$ utilized in the proposed model, which can be calculated based on the above known parameters. Their detailed derivation process is the same as what has been reported in the work of \cite{lagoudas2012,xu_2019sms}. Here they are provided to complete the model calibration process. Be noted that the values of $H^\textit{cur}$ and its partial derivative over stress in the following equations are calculated at the actual calibration test stress value. 

\begin{equation}\label{eq:Reduced_5_Parameters}
\begin{aligned}
\begin{cases}
&a_1=\rho_0 \Delta s_0 (M_f-M_s); \quad \\
&a_2=\rho_0 \Delta s_0 (A_s-A_f)\\
&a_3 = \cfrac{1}{4}~a_2(1+\cfrac{1}{n_3+1})-\cfrac{1}{4}~a_1(1+\cfrac{1}{n_1+1})\\
&\rho_0 \Delta u_0 =\dfrac{1}{2}\rho \Delta s_0 (M_s+A_f)\\
&Y_0=\dfrac{1}{2} \rho_0 \Delta s_0 (M_s-A_f)-a_3 \\[5pt]
&D = \dfrac{ (C_M-C_A)\big[ H^\textit{cur}+ \uptau\partial_{\uptau}H^\textit{cur}+ \uptau\Delta\mathcal{S}\big]+ +(C_M+C_A)(\Lambda^{tp}+\uptau\partial_{\uptau}\Lambda^{tp})}
{(C_M+C_A) ( H^\textit{cur} + \uptau\partial_{\uptau}H^\textit{cur})}\\[10pt]
&\rho_0 \Delta s_0 = -\dfrac{ 2 C_M C_A \big[ H^\textit{cur}+ \uptau\partial_{\uptau}H^\textit{cur}+ \uptau\Delta\mathcal{S}\big]}
{C_M+C_A} 
\end{cases}
\end{aligned}
\end{equation}

\end{document}